\documentclass{emulateapj}

\usepackage{times}
\usepackage{graphicx}
\usepackage{amssymb}
\usepackage{amsmath}
\usepackage{pifont}
\usepackage{graphics}
\usepackage{xcolor}
\usepackage{epsfig}
\usepackage{verbatim}
\usepackage{booktabs}

\newcommand{\feh}{\ensuremath{\left[{\rm Fe}/{\rm H}\right]}}  

\newcommand{\mstar}{\ensuremath{M_*}}
\newcommand{\rstar}{\ensuremath{R_*}}
\newcommand{\rplanet}{R\textsubscript{P}}          
\newcommand{\bjdtdb}{\ensuremath{\rm {BJD_{TDB}}}}
\newcommand{\teff}{\ensuremath{T_{\rm eff}}}
\newcommand{\teq}{\ensuremath{T_{\rm eq}}}
\newcommand{\loggstar}{\ensuremath{\log{g_*}}}
\newcommand{\loggplanet}{\ensuremath{\log{g_P}}}
\newcommand{\vsinistar}{\ensuremath{v\sin{I_*}}}
\newcommand{\msun}{\ensuremath{\,M_\Sun}}
\newcommand{\rsun}{\ensuremath{\,R_\Sun}}
\newcommand{\lsun}{\ensuremath{\,L_\Sun}}
\newcommand{\rhostar}{\ensuremath{\,\rho_*}}
\newcommand{\mj}{\ensuremath{\,M_{\rm J}}}
\newcommand{\rj}{\ensuremath{\,R_{\rm J}}}
\newcommand{\fave}{\langle F \rangle}
\newcommand{\fluxcgs}{\ensuremath{\rm 10^9 erg~s^{-1} cm^{-2}}}

\newcommand{\kms}{\ensuremath{\rm km\ s^{-1}}}

\begin{document}
\title{KELT-20\MakeLowercase{b}: A giant planet 
with a period of $P\sim 3.5$~days transiting the $V\sim 7.6$ early A star HD~185603}

\author{
Michael B. Lund\altaffilmark{1},
Joseph E. Rodriguez\altaffilmark{2},
George Zhou\altaffilmark{2},
B. Scott Gaudi\altaffilmark{3},
Keivan G. Stassun\altaffilmark{1,4},
Marshall C. Johnson\altaffilmark{3},
Allyson Bieryla\altaffilmark{2},
Ryan J. Oelkers\altaffilmark{1},
Daniel J. Stevens\altaffilmark{3},
Karen A. Collins\altaffilmark{1},
Kaloyan Penev\altaffilmark{5},
Samuel N. Quinn\altaffilmark{2},
David W. Latham\altaffilmark{2},
Steven Villanueva Jr.\altaffilmark{3},
Jason D. Eastman\altaffilmark{2},
John F. Kielkopf\altaffilmark{6},
Thomas E. Oberst\altaffilmark{7},
Eric L. N. Jensen\altaffilmark{8},
David H. Cohen\altaffilmark{8},
Michael D. Joner\altaffilmark{9},
Denise C. Stephens\altaffilmark{9},
Howard Relles\altaffilmark{2},
Giorgio Corfini\altaffilmark{10,*},
Joao Gregorio\altaffilmark{11},
Roberto Zambelli\altaffilmark{10},
Gilbert A. Esquerdo\altaffilmark{2},
Michael L. Calkins\altaffilmark{2},
Perry Berlind\altaffilmark{2},
David R. Ciardi\altaffilmark{12},
Courtney Dressing\altaffilmark{13,14},
Rahul Patel\altaffilmark{15},
Patrick Gagnon\altaffilmark{16,12},
Erica Gonzales\altaffilmark{17},
Thomas G. Beatty\altaffilmark{18,19},
Robert J. Siverd\altaffilmark{20},
Jonathan Labadie-Bartz\altaffilmark{21},
Rudolf B. Kuhn\altaffilmark{22,23},
Knicole D. Col\'{o}n\altaffilmark{24},
David James\altaffilmark{25},
Joshua Pepper\altaffilmark{21},
Benjamin J. Fulton\altaffilmark{26, 27},
Kim K. McLeod\altaffilmark{28},
Christopher Stockdale\altaffilmark{29},
Sebastiano Calchi Novati\altaffilmark{15,30},
D. L. DePoy\altaffilmark{31, 32},
Andrew Gould\altaffilmark{3},
Jennifer L. Marshall\altaffilmark{31, 32},
Mark Trueblood\altaffilmark{33},
Patricia Trueblood\altaffilmark{33}
}

\altaffiltext{1}{Department of Physics and Astronomy, Vanderbilt University, Nashville, TN 37235, USA; michael.b.lund@vanderbilt.edu}
\altaffiltext{2}{Harvard-Smithsonian Center for Astrophysics, Cambridge, MA 02138, USA}
\altaffiltext{3}{Department of Astronomy, The Ohio State University, Columbus, OH 43210, USA}
\altaffiltext{4}{Department of Physics, Fisk University, 1000 17th Avenue North, Nashville, TN 37208, USA}
\altaffiltext{5}{Department of Astrophysical Sciences, Princeton University, Princeton, NJ 08544, USA}
\altaffiltext{6}{Department of Physics and Astronomy, University of Louisville, Louisville, KY 40292, USA}
\altaffiltext{7}{Department of Physics, Westminster College, New Wilmington, PA 16172, USA}
\altaffiltext{8}{Department of Physics and Astronomy, Swarthmore College, Swarthmore, PA 19081, USA}
\altaffiltext{9}{Department of Physics and Astronomy, Brigham Young University, Provo, UT 84602, USA}
\altaffiltext{10}{Societ Astronomica Lunae, Castelnuovo Magra 19030, Italy}
\altaffiltext{11}{Atalaia Group \& CROW Observatory, Portalegre, Portugal}
\altaffiltext{12}{NASA Exoplanet Science Institute/Caltech, Pasadena, CA, USA}
\altaffiltext{13}{NASA Sagan Fellow, Division of Geological \& Planetary Sciences, California Institute of Technology, Pasadena, CA 91125, USA}
\altaffiltext{14}{Department of Astronomy, University of California, Berkeley, CA 94720-3411, USA}
\altaffiltext{15}{IPAC, Mail Code 100-22, Caltech, 1200 E. California Blvd., Pasadena, CA 91125, USA}
\altaffiltext{16}{College of the Canyons, 26455 Rockwell Canyon Rd., Santa Clarita, CA 91355, USA}
\altaffiltext{17}{University of California, Santa Cruz, CA, USA}
\altaffiltext{18}{Department of Astronomy \& Astrophysics, The Pennsylvania State University, 525 Davey Lab, University Park, PA 16802, USA}
\altaffiltext{19}{Center for Exoplanets and Habitable Worlds, The Pennsylvania State University, 525 Davey Lab, University Park, PA 16802, USA}
\altaffiltext{20}{Las Cumbres Observatory, 6740 Cortona Dr., Suite 102, Goleta, CA 93117, USA}
\altaffiltext{21}{Department of Physics, Lehigh University, 16 Memorial Drive East, Bethlehem, PA, 18015, USA}
\altaffiltext{22}{South African Astronomical Observatory, PO Box 9, Observatory, 7935 Cape Town, South Africa}
\altaffiltext{23}{Southern African Large Telescope, PO Box 9, Observatory, 7935 Cape Town, South Africa}
\altaffiltext{24}{NASA Goddard Space Flight Center, Greenbelt, MD 20771, USA}
\altaffiltext{25}{Astronomy Department, University of Washington, Box 351580, Seattle, WA 98195, USA}
\altaffiltext{26}{Institute for Astronomy, University of Hawaii, 2680 Woodlawn Drive, Honolulu, HI 96822-1839, USA}
\altaffiltext{27}{California Institute of Technology, Pasadena, CA 91125, USA}
\altaffiltext{28}{Department of Astronomy, Wellesley College, Wellesley, MA 02481, USA}
\altaffiltext{29}{Hazelwood Observatory, Churchill, Victoria, Australia}
\altaffiltext{30}{Dipartimento di Fisica “E. R. Caianiello”, Universit\'a di Salerno, Via Giovanni Paolo II 132, 84084 Fisciano (SA), Italy}
\altaffiltext{31}{George P. and Cynthia Woods Mitchell Institute for Fundamental Physics and Astronomy, Texas A \& M University, College Station, TX 77843, USA}
\altaffiltext{32}{Department of Physics and Astronomy, Texas A \& M University, College Station, TX 77843, USA}
\altaffiltext{33}{Winer Observatory, Sonoita, AZ 85637, USA}
\altaffiltext{*}{This paper is dedicated to the memory of Giorgio Corfini, who passed away in December 2014}

\shorttitle{KELT-20\MakeLowercase{b}}

\begin{abstract}
We report the discovery of KELT-20b, a hot Jupiter transiting a $V\sim 7.6$ early A star with an orbital period of $P\simeq 3.47$ days. We identified the initial transit signal in KELT-North survey data. Archival and follow-up photometry, the Gaia parallax, radial velocities, Doppler tomography, and adaptive optics imaging were used to confirm the planetary nature of the companion and characterize the system.
From global modeling we infer that the host star HD 185603 is a rapidly-rotating ($\vsinistar\simeq 120~\kms$) A2V star with an effective temperature of $\teff=8730^{+250}_{-260}$\,K, mass of $\mstar=1.76^{+0.14}_{-0.20}$ \msun, radius of 
$\rstar=1.561^{+0.058}_{-0.064}$ \rsun, surface gravity of $\loggstar=4.292^{+0.017}_{-0.020}$, and age of $\la 600$~Myr.
The planetary companion has a radius of $\rplanet=1.735^{+0.070}_{-0.075}~\rj$, a semimajor axis
of $a=0.0542^{+0.0014}_{-0.0021}$AU, and a linear ephemeris of
$\bjdtdb=2457503.120049 \pm 0.000190 + E(3.4741070\pm0.0000019)$. We place a $3\sigma$ upper limit of
$\sim 3.5~\mj$ on the mass of the planet. The Doppler tomographic measurement indicates that the planetary orbit normal is well aligned with the projected spin-axis of the star ($\lambda= 3.4\pm {2.1}$ degrees). The inclination of the star is constrained to be $24.4<I_*<155.6$ degrees, implying a true (three-dimensional) spin-orbit alignment of $1.3<\psi<69.8$ degrees.
The planet receives an insolation flux of
$\sim 8\times 10^9~{\rm erg~s^{-1}~cm^{-2}}$, implying
an equilibrium temperature of of $\sim 2250$\,K, assuming zero albedo and
complete heat redistribution. Due to the high stellar $\teff$, the planet also receives
an ultraviolet (wavelengths $d\le 91.2$~nm) insolation flux of $\sim 9.1\times 10^4~{\rm erg~s^{-1}~cm^{-2}}$, which may lead to significant ablation of
the planetary atmosphere.  
Together with WASP-33, Kepler-13 A, HAT-P-57, KELT-17, and KELT-9, KELT-20 is the sixth A star host of a transiting giant planet, and 
the third-brightest host (in $V$) of a transiting planet. The system is a slightly longer-period analog of the KELT-9 system.
\end{abstract}

\keywords{
planets and satellites: detection --
planets and satellites: gaseous planets --
stars: individual (HD 185603) --
techniques: photometric --
techniques: radial velocities --
methods: observational
}

\section{Introduction}     
\label{sec:Intro}
The first surveys for exoplanets, which primarily used the radial velocity method\footnote{While not the focus of this introduction, we would be remiss not to note the discovery of the planetary companions to the pulsar PSR1257+12 by \citet{Wolszczan:1992}.}, focused on sunlike (late F, G and early K) dwarf stars.  This was due to the fact that old stars with \teff below the Kraft break \citep{Kraft:1967} at $\teff \simeq 6250$\,K tend to be slowly rotating and have plentiful absorption lines, therefore enabling the sub-tens of meters per second precision that was expected to be needed to detect analogs of the planets in our solar system.   Stars cooler than early K also have plentiful lines, but are generally faint in the optical, where these initial surveys were carried out.  Given the high-resolution ($R\ga 50,000$) spectra needed to resolve the stellar spectral lines, high photon counts were difficult to acquire for cooler stars with the modest-aperture telescopes that were then available at the time.  

Of course, it came as a surprise when the first exoplanets discovered around main-sequence stars \citep{Campbell:1988,Latham:1989,Mayor:1995,Marcy:1996} did not resemble the planets in our solar system, and typically induced much higher radial velocity (RV) amplitudes than even our own giant planets.  Indeed, the Jupiter-like planetary companion to 51 Pegasi \citep{Mayor:1995}, which jump-started the field of exoplanets (despite not being the first exoplanet discovered), has such a short period that it creates a reflex RV amplitude on its host star of hundreds of meters per second.  It is the prototypical ``hot Jupiter'', a class of planets that are now known to orbit $\sim 0.5-1\%$ of stars \citep{Gould:2006, Howard:2012, Wright:2012}, but whose origins and characteristics remain important topics of study. 

Subsequent surveys for exoplanets, including those using the transit \citep{Winn:2010} and microlensing \citep{Gaudi:2012} methods, began to more fully explore the planet populations of lower-mass stars, and in particular around M dwarfs. The reasons for this are clear: RV, transit, and microlensing surveys are all more sensitive to planets orbiting low-mass stars (albeit for different reasons, see \citealt{Wright:2013}).  For potentially habitable planets, in particular, transit surveys have an enormous advantage over other detection methods when targeting low-mass stars \citep{Gould:2003}.  This advantage has since been dubbed the "small star opportunity", and has been one of the many reasons that the {\it Kepler} \citep{Borucki:2010} mission, as well as other ground-based surveys such as MEarth \citep{Nutzman:2008,Charbonneau:2009, Berta:2012} and TRAPPIST \citep{Gillon:2014}, have been so impactful.  

Indeed, in the over 25 years since the first confirmed exoplanets were discovered, the number of known exoplanets has increased dramatically, to almost 3500 confirmed exoplanets and an additional 2200 unconfirmed planet candidates\footnote{From https://exoplanetarchive.ipac.caltech.edu/, accessed July 3, 2017}. As the field of exoplanets has developed, there have been two broad goals: determining the overall demographics of exoplanets and how these demographics depend on the properties of the planets and their host stars, and finding individual exoplanets that can be characterized in detail, in particular their atmospheres. The primary techniques for characterizing exoplanet atmospheres are transits and direct imaging. 
The combination of transit photometry and radial velocity measurements can provide a planet's radius and mass and, by extension, its density and bulk composition. Beyond this, phase curves and spectroscopy of transits and eclipses can shed light on the atmospheric properties of the system. 
Although planet densities can be determined even for quite faint host stars, detailed spectra and phase curves benefit greatly from having host stars that are bright \citep{Seager:2010}.  Indeed, finding such bright transit hosts is one of the primary motivations of the Transiting Exoplanet Survey Satellite \citep{Ricker:2015}.

The Kilodegree Extremely Little Telescope Survey (KELT; \citealt{Pepper:2003,Pepper:2007,Pepper:2012}) was originally designed to find transiting hot Jupiters orbiting bright ($8\la V \la 10$) stars, precisely the targets best suited for follow-up and atmospheric characterization.  Nevertheless, the KELT survey did not start actively vetting targets until around 2011, by which point many ground-based transit surveys had discovered a number of transiting planets orbiting moderately bright stars \citep{Alonso:2004,McCullough:2005,Bakos:2007,CollierCameron:2007}.

Concurrently, while the overall picture of the demographics of planetary systems orbiting late F to M stars was starting to become clear, the properties of planetary systems orbiting more massive and hotter stars remained relatively murky.  This was largely because the workhorse planet detection technique, RVs, begins to have difficulties achieving precisions of better than a few hundred meters per second for stars above $\teff \simeq 6250$K, both because these stars have thin convective envelopes and so do not spin down with age due to magnetic braking, and because they have fewer spectral lines than cooler stars.  Although there were some RV surveys that targeted A and F stars, these did not result in many detections (e.g., \citealt{Galland:2005}). 

Another avenue to studying planets orbiting more massive stars was to survey "Retired A Stars" \citep{Johnson:2007}, giant stars whose progenitors were, ostensibly, A stars while on the main sequence.  However, the difficulty of inferring the mass of a giant star through its observable properties led some to question whether this sample of stars was, indeed, evolved from more massive progenitors, or simply solar mass-analogs \citep{Lloyd:2011}. Although (as demonstrated by the discovery announced in this paper) photometric transit surveys are certainly sensitive to hot Jupiters orbiting hotter and more massive main sequence stars, the conventional wisdom for many years was that a positive RV detection was required to confirm a transiting planet candidate.  

This perception began to change around nearly the same time for independent, but related reasons.  First, the discovery of WASP-33b \citep{CollierCameron2010}, demonstrated that a combination of Doppler tomography and a robust upper limit on the companion mass from RV can confirm a transiting planet.  Second, the use of statistical tools by the {\it Kepler} mission also relaxed the perception that RV confirmation was needed to validate a planet.  These changes, together with the somewhat fortuitous and accidental discovery of KELT-1b \citep{Siverd:2012}, led the KELT collaboration to pursue planets around more massive and hotter stars.  

To date, including the planet KELT-20b announced here, six transiting giant planet companions to main-sequence A stars are known: WASP-33, Kepler-13 A, HAT-P-57, KELT-17, and KELT-9.  A few additional companions to hot stars or remnants have been announced from the {\it Kepler} mission via transits, pulsation timing or Doppler beaming (e.g., \citealt{Ahlers:2015,Charpinet:2011,Silvotti:2007,Silvotti:2014}).  Finally, several directly-imaged planets orbiting young stars with $\teff \ga 7500$K have been announced \footnote{We note that the primary to the directly-imaged planetary system, HR 8799 \citep{Marois:2008}, is often referred to as an A star, but has an effective temperature that is on the border between an A9V and F0V star \citep{Pecaut:2013}, and properties that are more reminiscent of a $\lambda$ Boo star.}, the three hottest of which have very large uncertainties in the masses and radii of the planets due to the uncertain age of their parent stars, which may put them in the brown dwarf regime \citep{Carson:2013, Lafreniere:2011, Acke:2006}.  One of the advantages of discovering transiting planets orbiting bright stars is that it is possible to estimate the mass and radius of the host star to good precision (see Sec.~\ref{sec:StellarMass}). 

KELT-9b is an exemplar with regard to understanding exoplanet structure around hot stars, as it is both the brightest (V magnitude of 7.55) and hottest (10,170K) star known to host a transiting hot Jupiter, and provides an excellent opportunity to characterize a planet that is receiving an extreme amount of stellar radiation \citep{Gaudi:2017}. In this paper, we present the discovery and characterization of KELT-20b, a system that provides a comparison to KELT-9b of a hot Jupiter orbiting a very hot main sequence host star. In particular, KELT-20 is the third brightest star to host a transiting planet (in $V$), and the second brightest to host a hot Jupiter ($V=7.58$) as well as the second hottest host star ($\teff=8730$~K). KELT-20b is comparatively much cooler than KELT-9b, but at \teq $\sim 2260$~K is still one of the hottest exoplanets yet discovered.

\section{Discovery and Follow-Up Observations}
\label{sec:Obs}

\subsection{Discovery}
\label{sec:Discovery}
\begin{figure}
\centering 
\includegraphics[width=1.1\columnwidth, angle=0, trim = 0 2.6in 0 0]{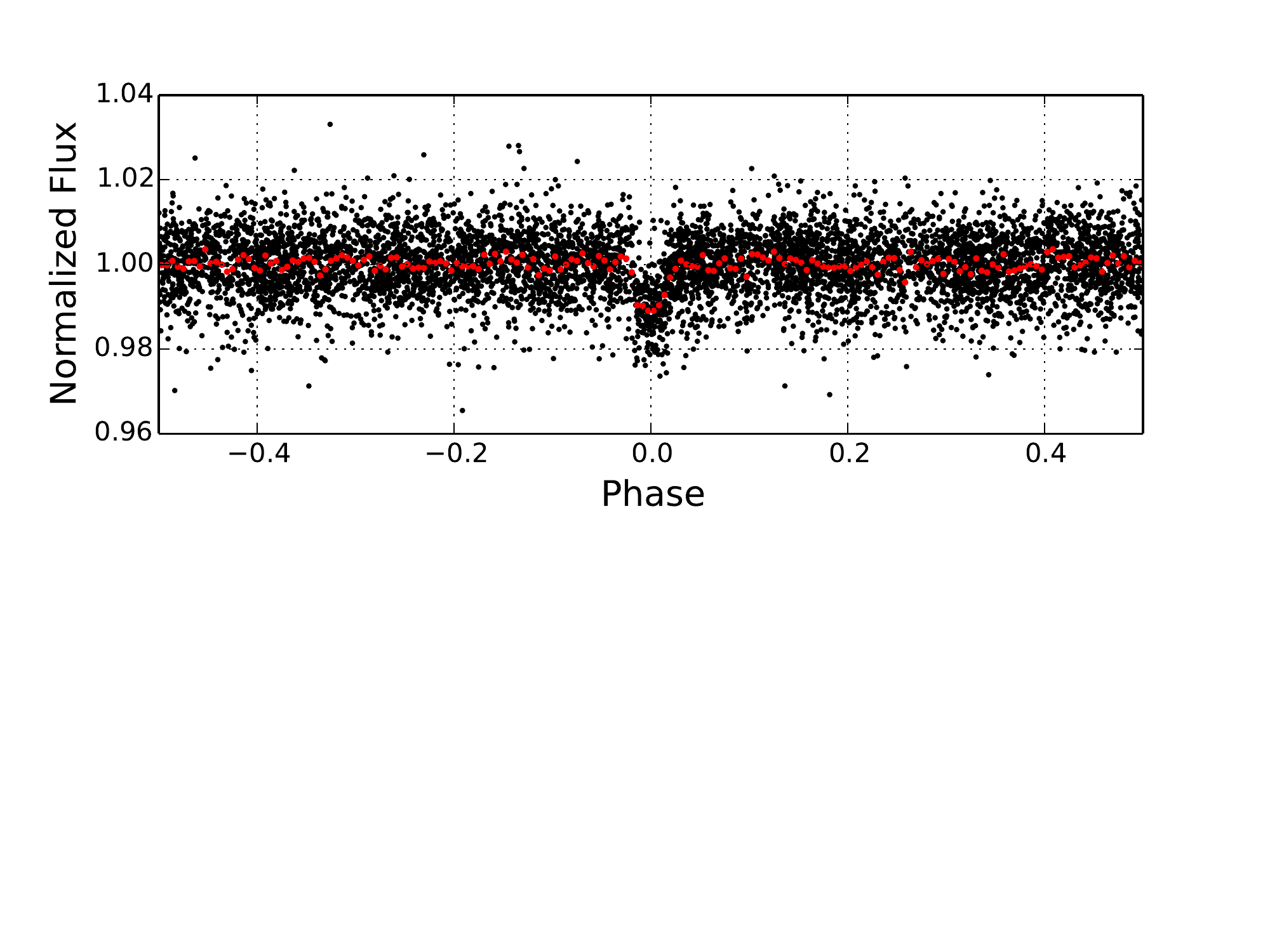}
\caption{\footnotesize Discovery light curve for KELT-20b based on 6740 observations from the KELT-North telescope.  The data have been phase-folded on the preliminary value for the period, 3.4739926\,d.}
\label{fig:DiscoveryLC}
\end{figure}

From a reduction of KELT-North field 11, KELT-20 (HD 185603) was identified as an exoplanet candidate following the same reduction and candidate selection process as described in detail in \citealt{Siverd:2012}. KELT-North field 11 is a 26$\degr$ $\times$ 26$\degr$ area of the sky centered on $\alpha =$ 19$^{h}$ 27$^{m}$ 00$^{s}$, $\delta =$ 31$\degr$ 39$\arcmin$ 56$\farcs$16 J2000 and was observed 6740 times from UT 2007 May 30 to UT 2014 November 25. From our periodicity search using the VARTOOLS \citep{Hartman:2016} implementation of Box-Least-Squares fitting \citep{Kovacs:2002}, KELT-20b was identified as a candidate with a 3.4739926 day period, 3.06 hour transit duration, and a 0.81\% transit depth. The phase-folded discovery light curve containing all 6740 points is shown in Figure \ref{fig:DiscoveryLC}. We note that KELT-20b was first identified as a candidate in a prior reduction of KELT-North field 11 using data that ended in UT 2013 June 14 ($\sim$700 fewer observations than are shown in Figure \ref{fig:DiscoveryLC}). The BLS results mentioned above are those of the initial discovery parameters.  See Table~\ref{tab:LitProps} for the photometric and kinematic properties of KELT-20 from the literature and this work. 

\begin{table}
\footnotesize
\centering
\caption{Literature Properties for KELT-20 }
\begin{tabular}{llcc}
  \hline
  \hline
Other IDs\dotfill         & \multicolumn{3}{l}{HD 185603} \\
                        & \multicolumn{3}{l}{TYC 2655-3344-1}				\\
	                   & \multicolumn{3}{l}{2MASS J19383872+3113091}
\\
\hline
Parameter & Description & Value & Ref. \\
\hline
$\alpha_{\rm J2000}$\dotfill	&Right Ascension (RA)\dotfill & $19^h38^m38\fs73$			& 1	\\
$\delta_{\rm J2000}$\dotfill	&Declination (Dec)\dotfill & +31\arcdeg13\arcmin09\farcs21			& 1	\\
\\
156.5~nm \dotfill	      &	USST (cW/m$^{2}$/nm/$10^{12}$)\dotfill  & $1.51\pm0.17$ & 2\\
196.5~nm \dotfill	      &	USST (cW/m$^{2}$/nm/$10^{12}$)\dotfill  & $3.30\pm0.24$ & 2\\
236.5~nm \dotfill	      &	USST (cW/m$^{2}$/nm/$10^{12}$)\dotfill  & $2.55\pm0.15$ & 2\\
274.0~nm \dotfill	      &	USST (cW/m$^{2}$/nm/$10^{12}$)\dotfill  & $2.27\pm0.07$ & 2\\
\\
$B_{\rm T}$\dotfill			&Tycho $B_{\rm T}$ mag.\dotfill & 7.697 $\pm$ 0.015		& 3	\\
$V_{\rm T}$\dotfill			&Tycho $V_{\rm T}$ mag.\dotfill & 7.592 $\pm$ 0.010		& 3	\\
\\
$u_{\rm Str}$\dotfill			& $u_{\rm Str{\ddot{o}}mgren-Crawford}$ mag.\dotfill & 9.094 $\pm$ 0.039		& 4	\\
$v_{\rm Str}$\dotfill			& $v_{\rm Str{\ddot{o}}mgren-Crawford}$ mag.\dotfill & 7.874 $\pm$ 0.024		& 4	\\
$b_{\rm Str}$\dotfill			& $b_{\rm Str{\ddot{o}}mgren-Crawford}$ mag.\dotfill & 7.645 $\pm$ 0.014		& 4	\\
$y_{\rm Str}$\dotfill			& $y_{\rm Str{\ddot{o}}mgren-Crawford}$ mag.\dotfill & 7.610 $\pm$ 0.010		& 4	\\
\\
$J$\dotfill		     	&   2MASS $J$ mag.\dotfill & 7.424  $\pm$ 0.024	         	& 5	\\
$H$\dotfill			    & 2MASS $H$ mag.\dotfill & 7.446 $\pm$ 0.018	            & 5	\\
$K_{\rm S}$\dotfill			& 2MASS $K_{\rm S}$ mag.\dotfill & 7.415 $\pm$ 0.017	& 5	\\
\\
\textit{WISE1}\dotfill		& \textit{WISE1} mag.\dotfill & 7.394 $\pm$ 0.027		& 6	\\
\textit{WISE2}\dotfill		& \textit{WISE2} mag.\dotfill & 7.437 $\pm$ 0.020		& 6 \\
\textit{WISE3}\dotfill		& \textit{WISE3} mag.\dotfill & 7.439 $\pm$ 0.016		& 6	\\
\textit{WISE4}\dotfill		& \textit{WISE4} mag.\dotfill & 7.350 $\pm$ 0.097		& 6	\\
\\
$\mu_{\alpha}$\dotfill		& Gaia DR1 proper motion\dotfill & 3.261 $\pm$ 0.026 		& 7 \\
                    & \hspace{3pt} in RA (mas yr$^{-1}$)	& & \\
$\mu_{\delta}$\dotfill		& Gaia DR1 proper motion\dotfill 	&  -6.041 $\pm$ 0.032 &  7 \\
                    & \hspace{3pt} in DEC (mas yr$^{-1}$) & & \\
\\
$RV$\dotfill & Systemic radial \hspace{9pt}\dotfill  & $-23.3 \pm 0.3$ & \S\ref{sec:Spectra} \\
     & \hspace{3pt} velocity (\kms)  & & \\
$v\sin{I_\star}$\dotfill &  Projected stellar rotational \hspace{7pt}\dotfill &  114.0$\pm$4.3 & \S\ref{sec:SpinOrbit} \\
                 & \hspace{3pt} velocity (\kms)  & & \\
Spec. Type\dotfill & Spectral Type\dotfill & A2V & \S\ref{sec:SED} \\ 
Age\dotfill & Age (Myr)\dotfill & $\la 600$ & \S\ref{sec:Evol} \\
$\pi$\dotfill & Gaia Parallax (mas) \dotfill & 7.41 $\pm$ 0.39 & 5\dag \\
$d$\dotfill& Gaia-inferred distance (pc) \dotfill & 139.7 $\pm 6.6$ & 5\dag \\
$A_V$\dotfill & Visual extinction (mag) & 0.07 $\pm$ 0.07 & \S\ref{sec:SED} \\
$\Theta$\dotfill & Angular Diameter (mas) & 0.0555 $\pm$ 0.0070 & \S\ref{sec:SED} \\
$U^{*}$\dotfill & Space motion (\kms)\dotfill & 1.13$\pm$ 0.17  & \S\ref{sec:UVW} \\
$V$\dotfill       & Space motion (\kms)\dotfill & -8.98 $\pm$ 0.27 & \S\ref{sec:UVW} \\
$W$\dotfill       & Space motion (\kms)\dotfill & 0.75 $\pm$ 0.18 & \S\ref{sec:UVW} \\
\hline
\hline
\end{tabular}
\begin{flushleft} 
 \footnotesize{ \textbf{\textsc{NOTES:}}
    References are: $^1$\citet{vanLeeuwen:2007},$^2$\citet{Thompson:1995},$^3$\citet{Hog:2000}, $^4$\citet{Paunzen:2015}, $^5$\citet{Cutri:2003}, $^6$\citet{Cutri:2012},$^7$\citet{Brown:2016} Gaia DR1 http://gea.esac.esa.int/archive/
    \dag Gaia parallax after correcting for the systematic offset of $-0.21$~mas as described in \citet{Stassun:2016}.
}
\end{flushleft}

\label{tab:LitProps}
\end{table}

\subsection{Photometric Follow-up from KELT-FUN}
\label{sec:Photom}
We obtained follow-up time-series photometry from the KELT Follow-Up Network (KELT-FUN) to better characterize the transit depth, duration, and shape, as well as to check for potential astrophysical false positives. We used a custom version of the TAPIR software package \citep{Jensen:2013} to predict transits, and we observed 13 transits in a variety of bands between August 2014 and June 2017, as listed in Table~\ref{tab:Photom}. In Figure~\ref{fig:All_light curve} we display the photometry from all KELT-FUN observations, as well as the transit light curve when all follow-up observations are combined. Unless otherwise stated, all data were calibrated and analyzed using the AstroImageJ package\footnote{http://www.astro.louisville.edu/software/astroimagej} \citep{Collins:2013, Collins:2017}.
\subsubsection{Peter van de Kamp Observatory (PvdK)} 
We observed KELT-20b from the Swarthmore College Peter van de Kamp Observatory (PvdK) on UT 2014 August 29 and UT 2017 May 08 in the i' band. The observations came from an 0.6 m RCOS telescope with an Apogee U16M 4K$\times$4K CCD, giving a 26$\arcmin$ $\times$ 26$\arcmin$ field of view. Using 2x2 binning, it has a pixel scale of 0$\farcs$76 pixel$^{-1}$.
\subsubsection{GCO} 
We observed KELT-20b from Giorgio Corfini's private observatory (GCO) in Lucca, Italy on UT 2014 September 25. The observations came from a 0.2 m Newtonian telescope with a SBIG STT-6303 ME CCD 1536$\times$1024 pixel camera, having a 59$\arcmin$ $\times$ 39$\arcmin$ field of view and a pixel scale of 2$\farcs$3 pixel$^{-1}$.
\subsubsection{WCO} 
We observed KELT-20b from the Westminster College Observatory (WCO) on UT 2015 October 06, UT 2017 May 08, and UT 2017 May 15 in the z' band. The observations came from a 0.35\,m f/11 Celestron C14 Schmidt-Cassegrain telescope and SBIG STL-6303E CCD with a ∼ 3k$\times$2k array of 9 $\mu$m pixels, having a 24$\arcmin$ $\times$ 16$\arcmin$ field of view and 1$\farcs$4 pixel$^{-1}$ image scale at 3 $\times$ 3 pixel binning.
\subsubsection{DEMONEXT} 
We observed KELT-20b using the DEMONEXT telescope \citep{Villanueva:2016} at Winer Observatory in Sonoita, AZ on UT 2016 May 21, UT 2016 June 04, and UT 2016 June 11 in the i' band. DEMONEXT is an 0.5\,m PlaneWave CDK20 f/6.8 Corrected Dall-Kirkham Astrograph telescope with a 2048$\times$2048 pixel FLI Proline CCD3041 camera, having a 30$\farcm$7 $\times$ 30$\farcm$7 field of view and a pixel scale of 0$\farcs$90 pixel$^{-1}$.
\subsubsection{MINERVA} 
We observed KELT-20b using one of the MINERVA project telescopes \citep{Swift:2015} on UT 2016 November 05. MINERVA consists of four 0.7\,m PlaneWave CDK-700 telescopes, located at the Fred L. Whipple Observatory on Mount Hopkins, AZ. A single MINERVA telescope has an Andor iKON-L 2048$\times$2048 camera, giving a field of view of 20$\farcm$9 $\times$ 20$\farcm$9, and a plate scale of 0$\farcs$6 pixel$^{-1}$. 
\subsubsection{MORC} 
We observed KELT-20b from Moore Observatory (MORC), operated by the University of Louisville, on UT 2017 May 08 in the i' band. The observations came from an 0.6\,m RCOS telescope with an Apogee U16M 4K$\times$4K CCD, giving it a 26$\arcmin$ $\times$ 26$\arcmin$ and 0$\farcs$39 pixel$^{-1}$.
\subsubsection{CDK20N} 
We observed KELT-20b from Moore Observatory (CDK20N), operated by the University of Louisville, on UT 2017 May 08 in the z' band. The observations came from an 0.5\,m Planewave Corrected Dall Kirkham telescope with an Apogee U16M 4K$\times$4K CCD, giving it a 37$\arcmin$ $\times$ 37$\arcmin$ field at 0$\farcs$54 pixel$^{-1}$.
\subsubsection{CROW} 
We observed KELT-20b from Canela’s Robotic Observatory (CROW) in Portalegre, Portugal on UT 2017 June 11 in the z' band. The observations came from an 0.3\,m Schmidt-Cassegrain telescope with a KAF-3200E CCD, having a 30$\arcmin$ $\times$ 20$\arcmin$ field of view and a pixel scale of 0$\farcs$84 pixel$^{-1}$.

\begin{figure}
\vspace{.0in}
\includegraphics[width=1\linewidth,height=5in]{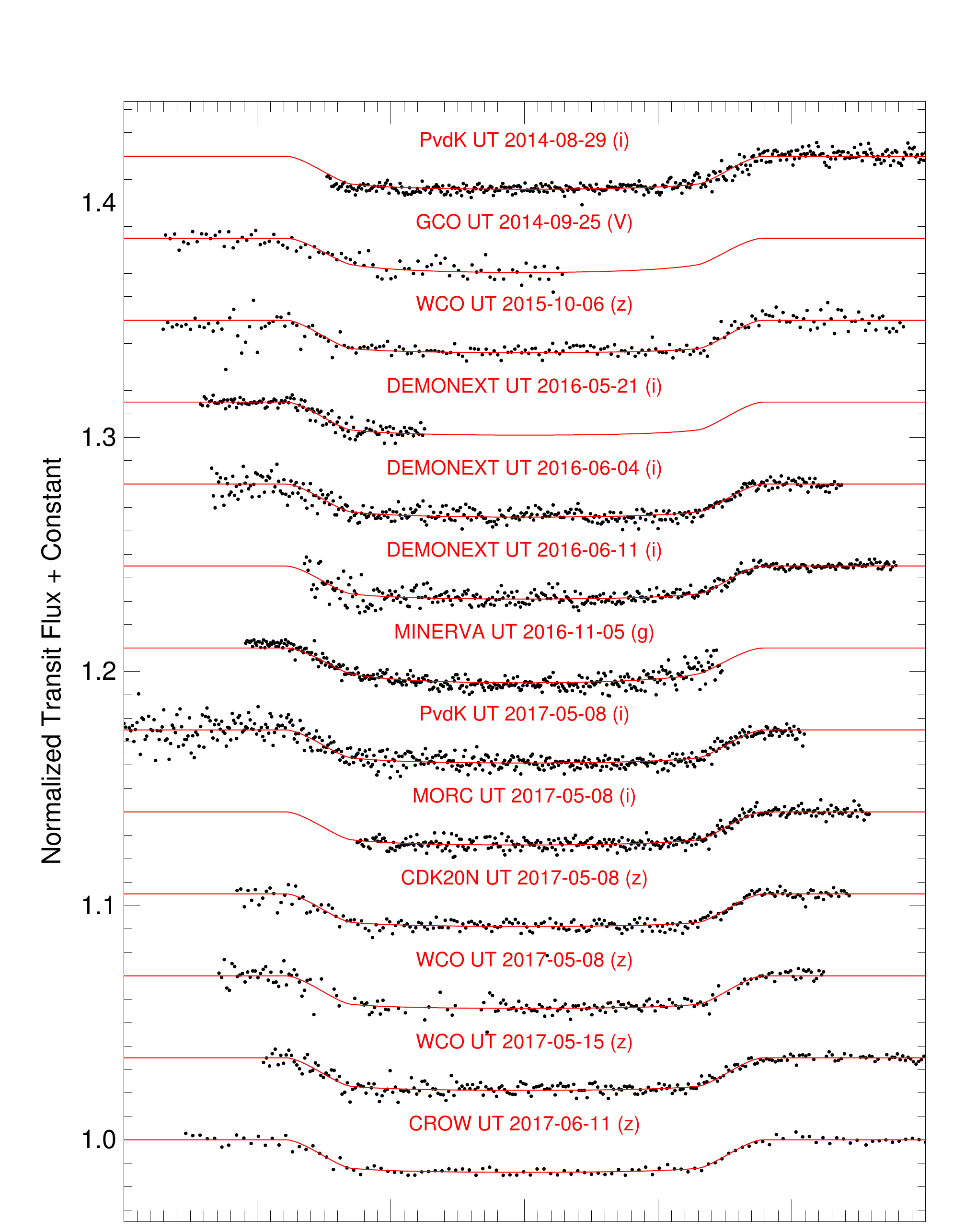}
\vspace{-.25in}

\includegraphics[width=1\linewidth]{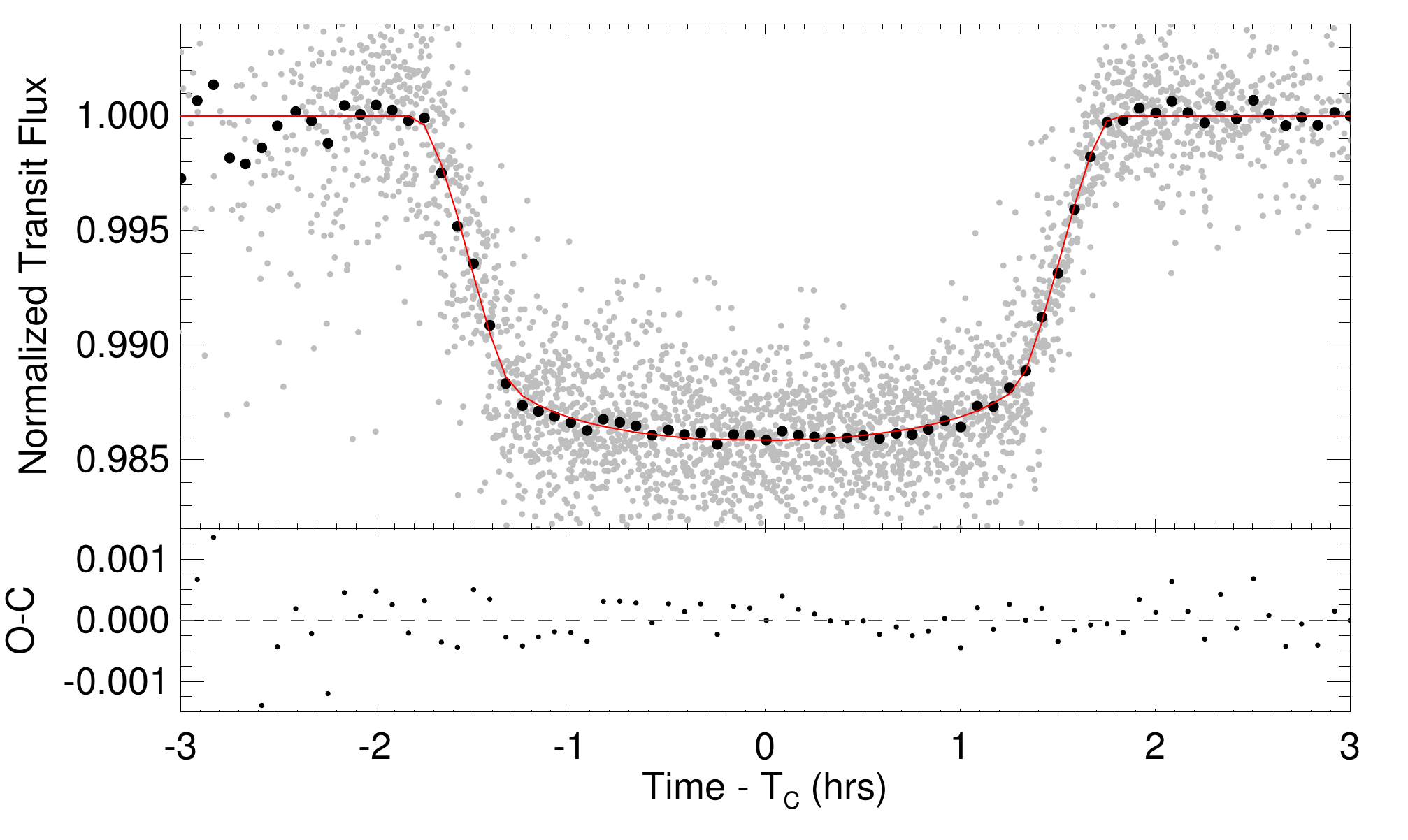}
\caption{(Top) The follow-up observations of KELT-20b by the KELT Follow-Up Network. The red line represents the best fit model for each transit. (Bottom) All follow-up transits combined into one light curve (grey) and a 5 minuted binned light curve (black). The red line is the combined and binned models for each transit.
}
\label{fig:All_light curve} 
\end{figure}

\begin{table*}
 \footnotesize
 \centering
 \setlength\tabcolsep{1.5pt}
 \caption{Photometric follow-up observations of KELT-20\textrm{\normalfont b}}
 \begin{tabular}{lcccccccc}
   \hline
   \hline

Observatory & Location & Aperture & Plate scale& Date      & Filter  & Exposure & Detrending parameters$^a$  \\
            &          & (m)      & ($\rm \arcsec~pix^{-1}$)& (UT) &   & Time (s) & &  \\
\hline

PvdK        & PA, USA       & 0.6      & 0.76       & 2014 Aug 29    & i$^\prime$      & 20       & airmass, time\\
GCO         & Lucca, Italy & 0.2      & 2.3       & 2014 Sept 25   & $V$            & 90         & airmass \\
WCO         & PA, USA       & 0.35     & 1.45       & 2015 Oct 06   & z$^\prime$      & 12       & airmass\\
DEMONEXT    & AZ, USA       & 0.5    &   0.90          & 2016 May 21   & i$^\prime$      & 31       & None\\
DEMONEXT    & AZ, USA       & 0.5    &   0.90          & 2016 June 04   & i$^\prime$      & 31       & None\\
DEMONEXT    & AZ, USA       & 0.5    &   0.90          & 2016 June 11   & i$^\prime$      & 31       & None\\
MINERVA     & AZ, USA       & 0.7    &   0.60          & 2016 Nov 05   & g$^\prime$      & 31       & airmass\\
PvdK        & PA, USA       & 0.6      & 0.76       & 2017 May 08    & i$^\prime$      & 20       & airmass\\
MORC        & KY, USA     & 0.6      & 0.39       & 2017 May 08    & i$^\prime$      & 20       & airmass\\
CDK20N      & KY, USA     & 0.5      & 0.54       & 2017 May 08    & z$^\prime$      & 60,40,30       & airmass\\
WCO         & PA, USA       & 0.35     & 1.45       & 2017 May 08   & z$^\prime$      & 12       & airmass\\
WCO         & PA, USA       & 0.35     & 1.45       & 2017 May 15   & z$^\prime$      & 12       & airmass\\
CROW        & Portalegre, Portugal     & 0.3     & 0.84       & 2017 June 11   & z$^\prime$      & 150       & airmass\\
\hline
\hline
\multicolumn{7}{l}{$^a$Photometric parameters allowed to vary in global fits as described in the text. }
\end{tabular}
\label{tab:Photom}
\end{table*}

\subsection{Spectroscopic Follow-up}
\label{sec:Spectra}

\begin{figure}
\includegraphics[width=1\linewidth]{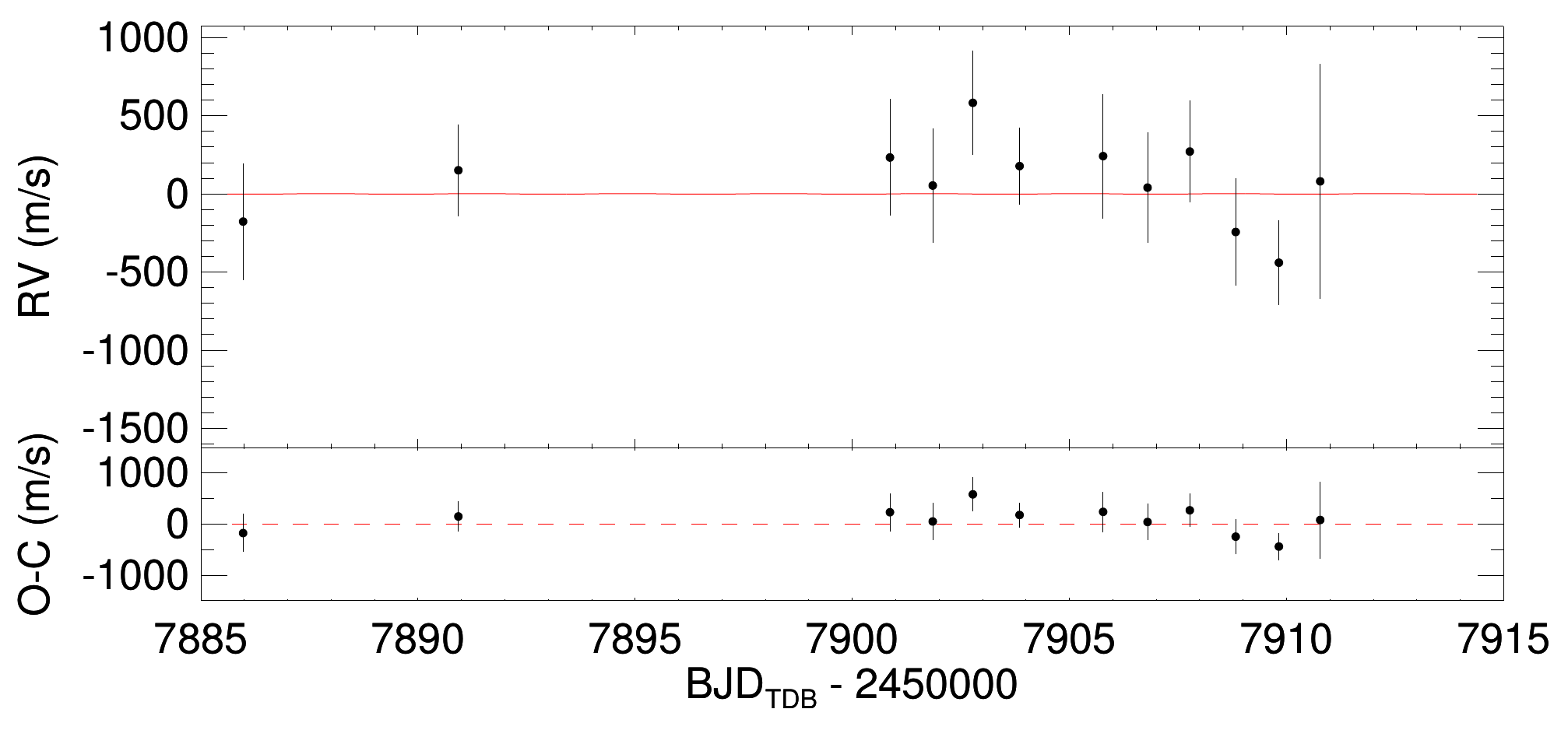}
   \vspace{-.3in}
\includegraphics[width=1\linewidth]{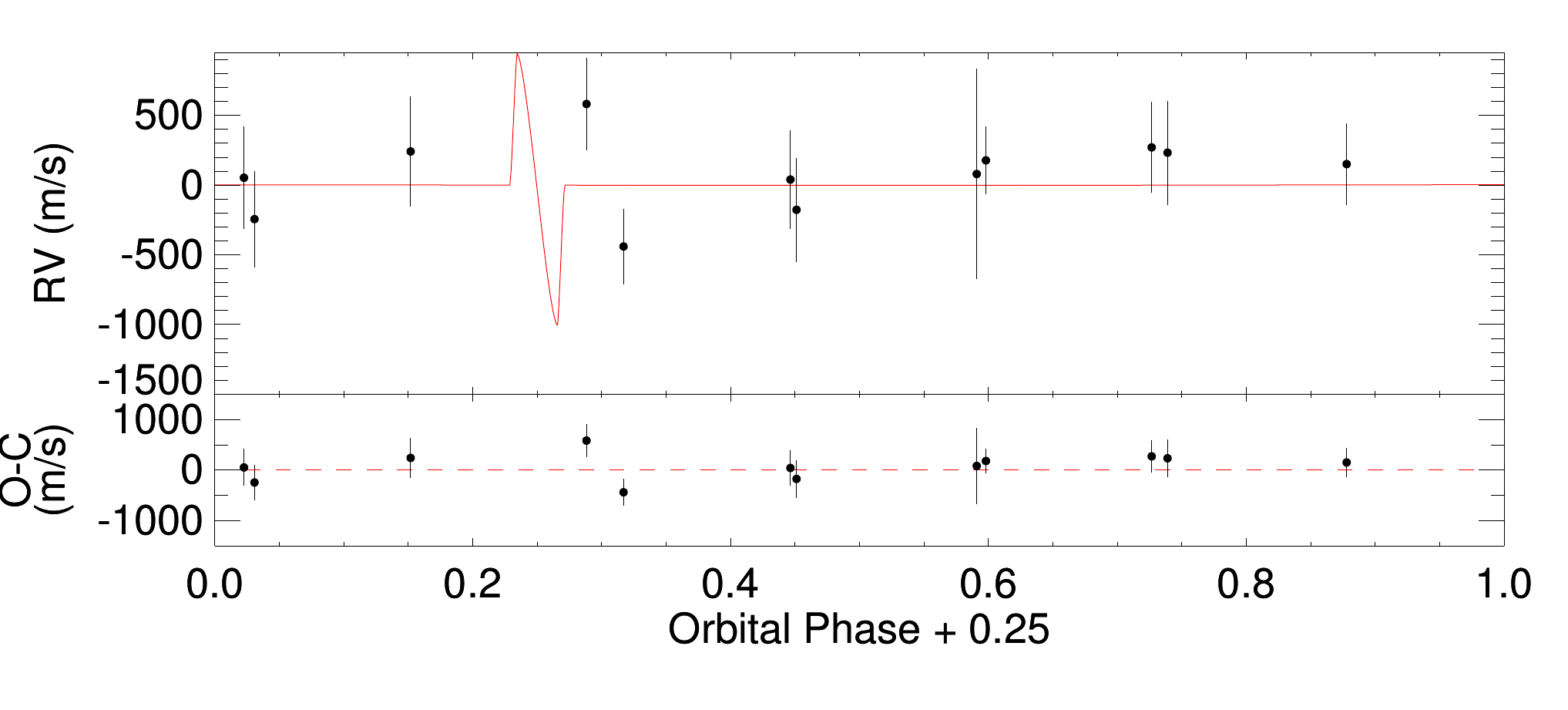}
   \vspace{.1in}
\caption{(Top) The TRES RV measurements of KELT-20b with the best fit model shown in red. The residuals to the fit are shown below. (Bottom) The RV measurements phase-folded to the global fit determined ephemeris. The predicted RM effect is shown at 0.25 phase. The residuals are shown below. }
\label{fig:RVs} 
\end{figure}

We obtained a series of spectroscopic follow-up observations of KELT-20b with the Tillinghast Reflector Echelle Spectrograph (TRES) on the 1.5 m telescope at the Fred Lawrence Whipple Observatory, Mount Hopkins, Arizona, USA. TRES is a fibre-fed echelle spectrograph, with a spectral resolution of $\lambda / \Delta \lambda \sim 44000$\, and a wavelength coverage of 3900 -- 9100\,\AA\  over the 51 orders. Radial velocites obtained over 11 out-of-transit orbital phases were used to constrain the mass of the planetary companion. Relative radial velocities were measured by cross correlating multiple orders of the TRES spectra against synthetic spectra and weight averaging the derived velocities, these `multi-order' velocities are listed in Table~\ref{tab:Spectra} and plotted in Figure~\ref{fig:RVs}.  In addition, 21 in-transit observations were obtained on the night of UT  2017-04-24 to measure the Doppler tomographic transit of the planet. The analysis of these observations is described in Section~\ref{sec:SpinOrbit}.

\begin{table}
\centering
 \caption{Relative RVs for KELT-20 from TRES}
 \label{tab:Spectra}
 \begin{tabular}{lrrrrl}
    \hline
    \hline
    \multicolumn{1}{l}{\bjdtdb} & \multicolumn{1}{c}{RV} 	& \multicolumn{1}{c}{$\sigma_{\rm RV}$}\\
    & \multicolumn{1}{c}{($\rm m~s^{-1}$)} &\multicolumn{1}{c}{($\rm m~s^{-1}$)} \\
    \hline                                                                                       
2457885.970564 & 0 & 397.81\\
2457890.927060 & 328.01 & 313.63\\
2457900.866772 & 409.74 & 397.81\\
2457901.852101 & 230.49 & 390.78\\
2457902.775118 & 759.53 & 355.68\\
2457903.851423 & 354.69 & 261.69\\
2457905.775362 & 418.55 & 424.11\\
2457906.798723 & 217.00 & 377.87\\
2457907.772196 & 447.92 & 347.19\\
2457908.828699 & -66.87 & 367.83\\
2457909.823202 & -263.25 & 287.26\\
2457910.774902 & 257.09 & 802.67\\
\hline
 \end{tabular}
 \begin{flushleft}
  \footnotesize{ \textbf{\textsc{NOTES:}} The TRES RV zeropoint is arbitrarily set to the first TRES value.
  }
\end{flushleft}
\end{table} 
\subsection{High Contrast AO Imaging}
\label{sec:AO}
We obtained high-resolution imaging for KELT-20 with the infrared camera PHARO behind the adaptive optics (AO) system P3K on the Palomar 200-inch Hale telescope.  PHARO has a pixel scale of $0\farcs025$ pixel$^{-1}$ \citep{Hayward:2001}, and the data were obtained in the narrow-band filter Br-$\gamma$ on UT 2017 May 05.

The AO data were obtained in a 5-point quincunx dither pattern with each dither position separated by 5$\arcsec$.  Each dither position was observed 3 times, each offset from the previous image by $1\arcsec$ for a total of 15 frames; the integration time per frame was 45 seconds. We use the dithered images to remove sky background and dark current, and then align, flat-field, and stack the individual images. The PHARO AO data have a resolution of $0\farcs09$ (FWHM).

\begin{figure}[!ht]
\includegraphics[width=0.95\linewidth]{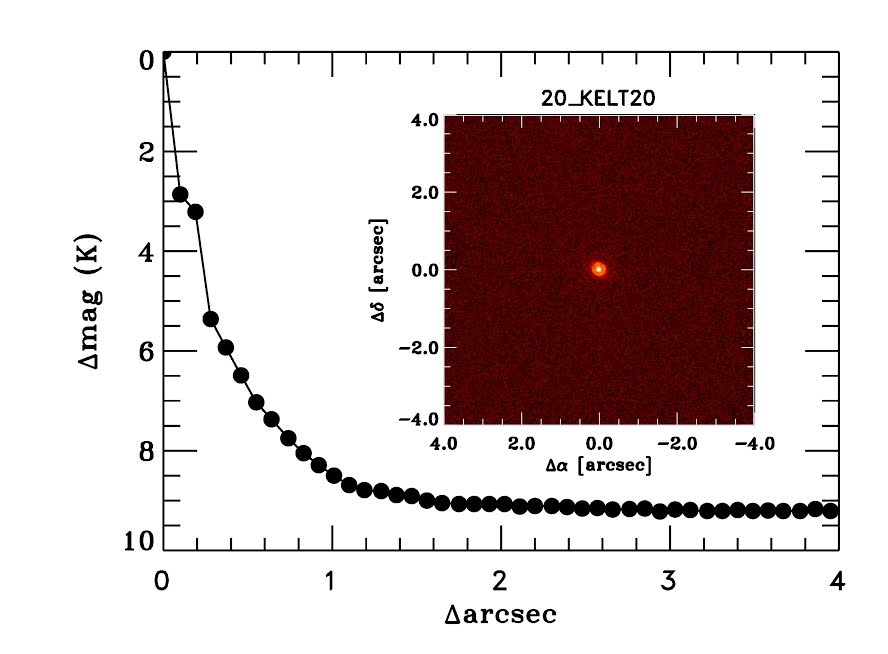}
\caption{The 5$\sigma$ contrast limit around KELT-20 in the PHARO AO data. \emph{Inset}: PHARO AO image of KELT-20.}
\label{fig:AO}
\end{figure}

The sensitivity of the AO data was determined by injecting simulated sources into the final combined images with separations from the primary targets in integer multiples of the central source's FWHM \citep{Furlan:2017}.  The sensitivity curve shown in Figure~\ref{fig:AO} represents the 5$\sigma$ limits of the imaging data.

For KELT-20, no stellar companions were detected in the infrared adaptive optics, indicating (to the limits of the data) that the star has no additional components to either dilute the transit depth or confuse the determination of the origin of the transit signal (e.g., \citet{Ciardi:2015}).

\section{Host Star Characterization}
\label{sec:Star}

\subsection{SED Analysis}\label{sec:SED}
\begin{figure}[!ht]
\includegraphics[width=1.0\linewidth, trim = 90 0 80 0]{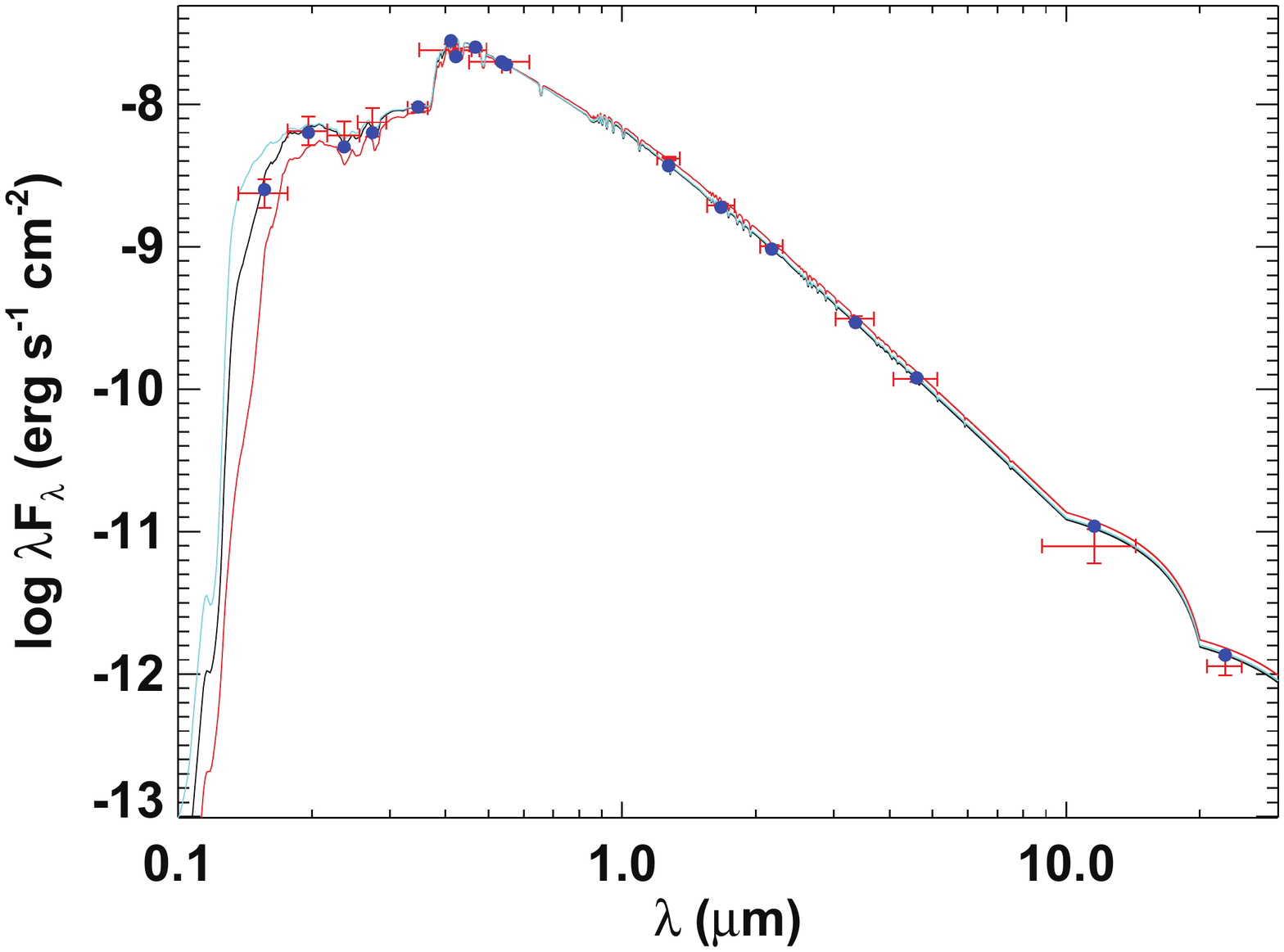}
\caption{\footnotesize Spectral energy distribution of KELT-20. The red crosses show observed broadband flux measurements, with vertical errorbars representing $1\sigma$ measurement uncertainty and horizontal errorbars representing the width of each bandpass. The blue dots are the predicted passband-integrated fluxes of the best-fit theoretical SED corresponding to our observed photometric bands. The best fit Kurucz atmosphere model is shown in black; the model atmospheres representing $\pm 1\sigma$ parameters are represented in cyan and red, respectively.}
\label{fig:SED}
\end{figure}

We assembled the available broadband photometry of KELT-20 (see Table~\ref{tab:LitProps}) in order to construct a spectral energy distribution (SED) spanning a large range of wavelengths from $\sim$0.15~\micron\ to 22~\micron\ (Figure~\ref{fig:SED}). We fit the SED using the model atmospheres of \citet{Kurucz:1992}, the free parameters being the stellar effective temperature (\teff), extinction ($A_V$), and a flux normalization factor (effectively the ratio of the stellar radius to the distance). The stellar surface gravity (\loggstar) and metallicity (\feh) have only a minor effect on the SED and are poorly constrained by this type of fit, so we simply adopted a solar metallicity and \loggstar = 4.3 (corroborated by the final global fit; see Sec.~\ref{sec:GlobalFit} and Table~\ref{tbl:KELT-17b}). The extinction was limited to the maximum value from the dust maps of \citet{Schlegel:1998} for this line of sight, $A_V$ = 1.43~mag.

The resulting best-fit parameters are $A_V = 0.07 \pm 0.07$ mag and \teff\/ = 8800$\pm$500 K, with a reduced chi-square of $\chi_\nu^2 = 3.05$ (Figure~\ref{fig:SED}). By directly integrating the (unextincted) fitted SED model, we obtain a semi-empirical measure of the stellar bolometric flux at Earth, $F_{\rm bol} = 2.46 \pm 0.27 \times 10^{-8}$ erg~s$^{-1}$~cm$^{-2}$. From $F_{\rm bol}$ and \teff\ we obtain a measure of the stellar angular radius, $\Theta$, which in turn provides a constraint on the stellar radius via the distance from the {\it Gaia} parallax of $R_\star = 1.61 \pm 0.22$~\rsun. This estimate of $R_\star$ is used as a constraint in the global system fit below (Sec.~\ref{sec:GlobalFit}). The \teff\ of 8800K corresponds to an A2V type star \citep{Pecaut:2013}.

\subsection{Nearly Empirical Estimate of the Stellar Mass}
\label{sec:StellarMass}

As was originally demonstrated in the context of transiting planets by \citet{Seager:2003}, under the assumption that $k\equiv R_P/R_*\ll 1$, it is possible to estimate the density ($\rhostar$) of a host star via a measurement of the full-width half-max (T\textsubscript{FWHM}) of the transit, the period ($P$), the impact parameter ($b$), the eccentricity and argument of periastron.  As these quantities can be measured essentially directly (i.e., without reliance on models), one can obtain an empirical estimate of $\rhostar$.  This can then be combined with the essentially direct estimate of $R_*$ as determined from $\teff$, the bolometric flux, and parallax above to estimate the stellar mass ($\mstar$), again without reliance on theoretical models (e.g., isochrones) or externally-calibrated relations (e.g., \citealt{Torres:2010}). This technique was recently applied to all transiting planets in the first Gaia data release by \citet{Stassun:2017}.

We do not have a constraint on the eccentricity or argument of periastron, but given the short period, it is reasonable to assume that the orbit has been circularized.  In the limit $e=0$ and $k\ll 1$,
\begin{equation}
M_* = \left(\frac{4P\rstar^3}{\pi G T_{FWHM}^3}\right)(1-b^2)^{\frac{3}{2}}.
\end{equation}
We adopt the estimates of $P$, T\textsubscript{FWHM}, and $b$ derived from global modeling (see Sec.~\ref{sec:GlobalFit}) using the Yale-Yonsei (YY) isochrone-constrained circular fits given in Tables \ref{tbl:KELT-17b} and \ref{tbl:KELT-17b_part2}.  We note that while these parameters formally rely on the constraints from the YY isochrones, since they are derived (almost) directly from data, their measurements are not, in fact, affected by these constraints. This can be seen by comparing the values of these parameters measured from the global modeling using the YY isochrones with those from the global modeling using the Torres relations; these parameters differ by $<1\%$ between these two fits in all cases.  Adopting the Gaia-inferred radius of $R_\star = 1.61 \pm 0.22$~\rsun, we find $M_*=1.90 \pm 0.47$~\msun, with an uncertainty of $\sim 25\%$.  We note that this uncertainty is dominated by the uncertainty in $\rstar$.  

Interestingly, this inferred mass is nearly identical to the mass inferred from the Torres-constrained global fit, and indeed the radius inferred from this global fit is nearly identical to the Gaia-determined radius. However, in both cases the uncertainties are somewhat smaller.  This implies that the mass and radius of the host are largely determined by the direct (model-independent) constraints in the Torres-constrained global fits, and completely consistent with the Torres relations.  The Torres relations are therefore primarily serving to decrease the uncertainties (slightly). 

Importantly, the inferred $\loggstar \simeq 4.3$ is at the higher end of what is typically expected from A stars of this $\teff$ and solar metallicity (see, e.g., \citealt{Torres:2010}).  This implies that the host is exceptionally close to (and perhaps lower than) the Zero Age Main Sequence (ZAMS) for solar-metallicity stars in the parameter space of $\loggstar$ versus $\teff$.  This can be explained in several ways.  First, the star could indeed have nearly solar metallicity, but be very young.  Second, the star could be older, but have sub-solar metallicity, since the ZAMS is at a lower $\loggstar$ at fixed $\teff$ for stars of lower metallicity.  Finally, the measurement of $\rstar$ from the SED and parallax could have a small systematic error.

Since the Torres relations do not encode age, it is possible for this star to have a higher $\loggstar$ at solar metallicity without resulting in any tension with the empirical parameters using those relations.  On the other hand, the YY isochrones do encode age, thus enforcing a maximum $\loggstar$ for a given metallicity (i.e., that of the ZAMS), and thus the inferred high $\loggstar$ disfavors this star having solar metallicity.  The YY isochrone fits therefore `prefer' lower metallicities for the host star, although we note that a solar metallicity is still allowed within $\sim 1\sigma$.  The lower metallicity inferred by the YY fits also results in a somewhat smaller mass and radius than inferred from the empirical methods above and the Torres-constrained global fits. 

Overall, we are agnostic about which of these three explanations are correct.  Generally, we note that A stars with metallicities of $\feh\sim-0.3$ are not common, and we note that the kinematics of this star (i.e., the low UVW velocities) support the interpretation the star is young.  Of course, we cannot rule out the simpler explanation that there are unrecognized subtle systematics affecting our inference of the radius, mass, and surface gravity of the star.

We note that a Hipparcos parallax also exists for this star, and is $8.73\pm 0.50$~mas. The radius and mass inferred from the Hipparcos parallax is $\rstar=1.37\pm 0.09$~\rsun and $\mstar=1.17\pm0.23$~\msun. These stellar parameters are inconsistent with those inferred from the Gaia parallax of $7.716 \pm 0.37$~mas.  In particular, as can be seen in Figures~\ref{fig:SED} and \ref{fig:hrd}, these values are completely inconsistent with the spectral energy distribution ($\teff$) or even the color of the source.  We therefore reject it and adopt the Gaia parallax with the \citet{Stassun:2016} systematic correction.
An examination of the reasons for this apparent discrepancy with the {\it Hipparcos} parallax is beyond the scope of this paper. Here we simply note the discrepancy and proceed with our analysis utilizing the {\it Gaia} parallax as a constraint on the system global solution (Sec.~\ref{sec:GlobalFit}).

\subsection{Evolutionary Analysis}
\label{sec:Evol}

\begin{figure}[!ht]
\centering
\includegraphics[width=0.75\columnwidth,angle=90]{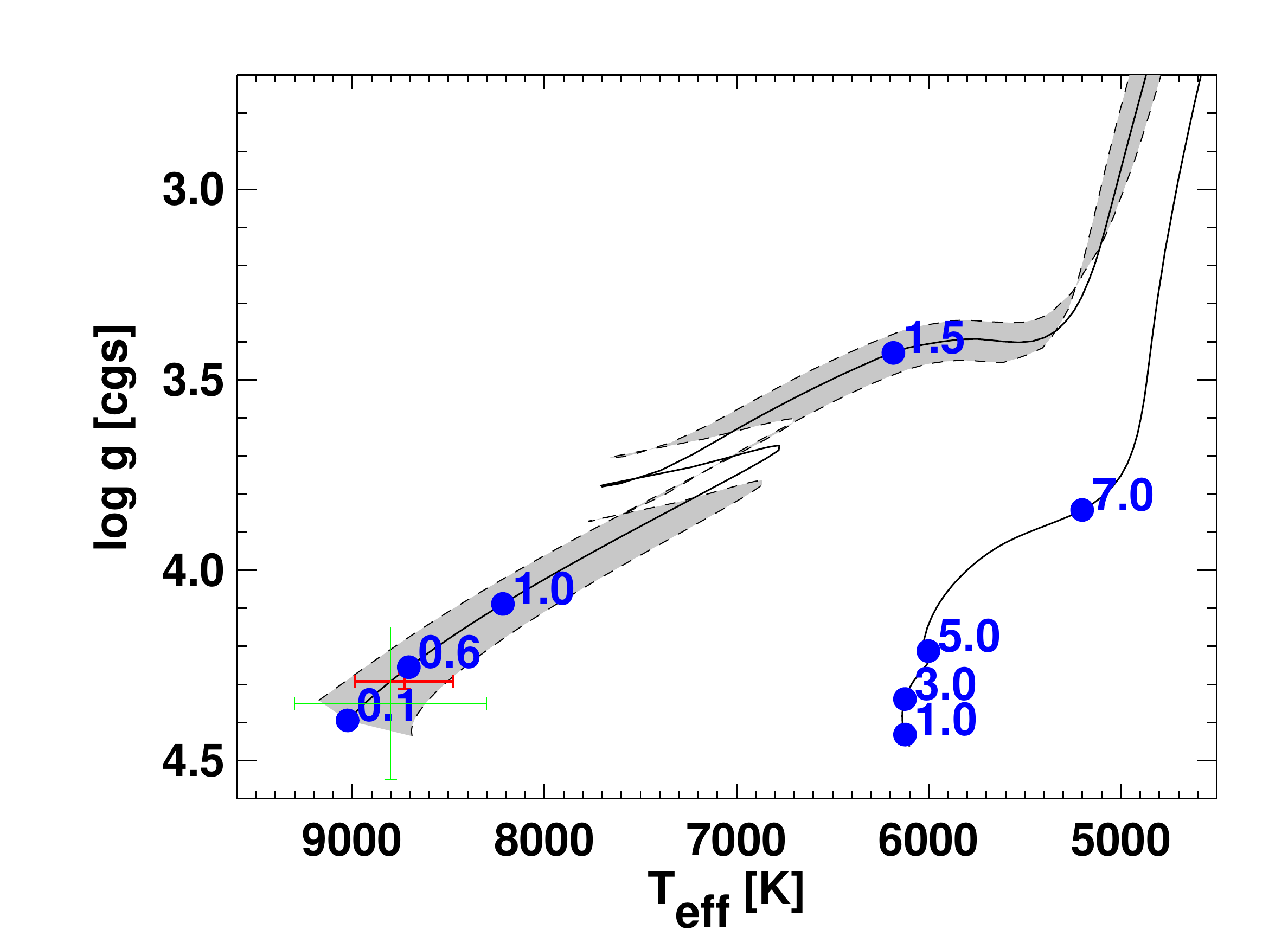}
\caption{KELT-20 in the modified Hertzsprung-Russell Diagram (\loggstar\ vs.\ \teff). The grey swath represents the Yonsei-Yale evolutionary track for a star with the mass inferred from the stellar radius (via the {\it Gaia} parallax and transit (see Sec.~\ref{sec:SED})), and $1\sigma$ error on that mass. Stellar ages (in Gyrs) along the evolutionary track are indicated with blue points. The  initial \teff\ and \loggstar\ inferred from the SED fit are represented by the green error bars; the final \teff\ and \loggstar\ from the global solution are represented by red error bars. For comparison, the evolutionary track for a star with the mass inferred from the {\it Hipparcos} parallax is also shown (see the text), and starts at a much cooler temperature.}
\label{fig:hrd}
\end{figure}

To put the KELT-20 system in context and to provide an initial estimate of the system age, we show in Figure~\ref{fig:hrd} the KELT-20 host star in the modified Hertzsprung-Russell diagram (\loggstar\ vs.\ \teff). Using the Yonsei-Yale stellar evolutionary models for a star of mass 1.76~\msun, we infer an age for KELT-20 of at most $\sim$600~Myr.

\subsection{Distance Above the Galactic Plane and UVW Space Motion}
\label{sec:UVW}

KELT-20 is located at equatorial coordinates $\alpha=19^h38^m38\fs73$,
and $\delta=+31\arcdeg13\arcmin09\farcs21$ (J2000), corresponding to Galactic
coordinates of $\ell=65.8\arcdeg$ and $b=4.6\arcdeg$.  Given the Gaia distance
of $139.7 \pm 6.6~{\rm pc}$ \citep{Brown:2016}, KELT-20 lies at a Galactocentric distance of roughly $8.26$~kpc, assuming a distance from the Sun to the Galactic center of $R_0=8.32$~kpc \citep{Gillenssen:2017}.  KELT-20 is located $\sim 10$~pc above the plane, well within the Galactic scale height for A stars of $\sim 50$~pc \citep{Bovy:2017}. 

Using the Gaia proper motion of $(\mu_\alpha,\mu_\delta)=(3.261 \pm 0.026, -6.041 \pm 0.032)~{\rm mas~yr}^{-1}$, the Gaia parallax, and the absolute radial velocity as determined from the TRES spectroscopy of $-23.8 \pm 0.3~{\rm km~s^{-1}}$, we find
that KELT-20 has a three-dimensional Galactic space motion of $(U,V,W)= (1.14 \pm 0.17, -8.98 \pm 0.27, 0.75 \pm 0.18)~{\rm km~s^{-1}}$, where positive $U$ is in the direction of the Galactic center, and we have adopted the \citet{Coskunoglu:2011} determination of the solar motion with respect to the local standard of rest. These values yield a 99.5\% probability that KELT-20 is a thin disk star, according to the classification scheme of \citet{Bensby:2003}, as expected for its young age and early spectral type.  

KELT-20 is projected against a supernova remnant, which is also visible in optical and H$\alpha$ survey data.  This is a known supernova remnant, SNR G065.3+05.7, which is about 0.8 kpc away (Boumis et al. 2004).  At a distance from {\it Gaia} of $\sim$140~pc, this is evidently a chance projection, with KELT-20 well in front of the supernova remnant.  

The line of sight toward KELT-20 in Cygnus is along the so-called Orion Spur or Orion Arm, and thus it would be expected that there would be a large population of young stars in that general direction.  Most of the young associations catalogued in that direction (e.g., the Cygnus OB associations, the North America Nebula, the Pelican Nebula, NGC 6914) lie at distances of 1~kpc or more, and we were not able to locate in the literature any evidence of known star-forming regions in the vicinity of the $\sim$140~pc distance to KELT-20.
We also checked KELT-20's Galactic space motion against the known young moving groups, and there is no obvious match. In addition, searching {\it Gaia} DR-1, there are no sources within 5 degrees of KELT-20 with similar proper motion and distance.  

Thus, while we cannot associate KELT-20 with any known star-forming region or known young stellar population in particular, its young age is completely plausible given its location in the Galaxy.  We infer that it was likely associated with some earlier episode of star formation in our spiral arm, but its local gas and any associated young stars have since dispersed into the field population.

\section{Planet Characterization}
\label{sec:Planet}

\subsection{EXOFAST Global Fit}
\label{sec:GlobalFit}

Using a heavily modified version of EXOFAST \citep{Eastman:2013}, an IDL-based exoplanet fitting suite, we perform a series of global fits to determine the system parameters for KELT-20. Within the global fit, all photometric and spectroscopic observations (including the Doppler tomography signal) are simultaneously fit. EXOFAST uses either the Yonsei-Yale (YY) stellar evolution model tracks \citep{Demarque:2004} or the Torres relations \citep{Torres:2010} to constrain the mass and radius of the host star, KELT-20. See \citet{Siverd:2012} for a detailed description of the global modeling routine. 

Within the global fit, each follow-up raw light curve and the determined detrending parameters shown in Table \ref{tab:Photom} are used as inputs for the fit. We impose a prior on \teff\ of 8800$\pm$500K determined from our SED analysis. Additionally, we are unable to precisely determine the metallicity of KELT-20 from our current observations, and so we set a prior on \feh\ of 0.0$\pm$0.5 dex. Further, we ran an initial global fit where a prior was set on the period and transit center time from our analysis of the KELT-North light curve. From performing a linear fit to the determined transit center times, we independently determined an ephemeris for KELT-20b (See \S\ref{sec:TTVs}). We reran the Torres and YY circular fits with a prior on the transit center time and period determined from this analysis. The KELT-North light curve is not included in any of the global fits we conducted. Lastly, we use the Gaia parallax shown in Table \ref{tab:LitProps} combined with the determined bolometric flux from our SED analysis to impose a prior on the host star's radius (R$_{\star}$ = 1.610$\pm$0.216). We perform two separate global fits where we fix the eccentricity of the planet's orbit to zero. One fit uses the YY models while the other uses the Torres relations to determine the mass and radius of KELT-20. For the discussion and interpretation of the KELT-20 system, we adopt the circular YY fit. The results of both fits are show in Tables~\ref{tbl:KELT-17b} and \ref{tbl:KELT-17b_part2}.

For the output parameters shown through this paper that use solar or Jovian units, we adopt the following constants throughout: {\it G\msun} = 1.3271244 $\times$ 10$^{20}$ m$^3$ s$^{-2}$, \rsun = 6.9566 $\times$ 10$^{8}$ m, \mj = 0.000954638698 \msun, and  \rj = 0.102792236 \rsun~ \citep{Standish:1995, Torres:2010, Eastman:2013, Prsa:2016}. 

\begin{table*}
 \scriptsize
\centering
\setlength\tabcolsep{1.5pt}
\caption{Median values and 68\% confidence interval for the physical and orbital parameters of the KELT-20 system}
  \label{tbl:KELT-17b}
  \begin{tabular}{lccccc}
  \hline
  \hline
  Parameter & Description (Units) & \textbf{Adopted Value} & Value  \\
  & & \textbf{(YY circular)} & (Torres circular) \\
 \hline
Stellar Parameters & & & \\
                               ~~~$M_{*}$\dotfill &Mass (\msun)\dotfill &$1.76_{-0.20}^{+0.14}$&$1.90_{-0.19}^{+0.21}$\\
                             ~~~$R_{*}$\dotfill &Radius (\rsun)\dotfill &$1.561_{-0.064}^{+0.058}$&$1.605\pm0.064$\\
                         ~~~$L_{*}$\dotfill &Luminosity (\lsun)\dotfill &$12.6_{-1.9}^{+2.2}$&$13.2_{-2.1}^{+2.3}$\\
                             ~~~$\rho_*$\dotfill &Density (cgs)\dotfill &$0.645_{-0.034}^{+0.036}$&$0.650_{-0.034}^{+0.037}$\\
                  ~~~$\log{g_*}$\dotfill &Surface gravity (cgs)\dotfill &$4.292_{-0.020}^{+0.017}$&$4.307\pm0.022$\\
                  ~~~$\teff$\dotfill &Effective temperature (K)\dotfill &$8730_{-260}^{+250}$&$8700_{-280}^{+260}$\\
                                 ~~~$\feh$\dotfill &Metallicity\dotfill &$-0.29_{-0.36}^{+0.22}$&$-0.02_{-0.48}^{+0.51}$\\
             ~~~$v\sin{I_*}$\dotfill &Rotational velocity (m/s)\dotfill &$115900\pm3400$&$115800\pm3400$\\
           ~~~$\lambda$\dotfill &Spin-orbit alignment (degrees)\dotfill &$3.4\pm2.1$&$3.4\pm2.1$\\
         ~~~$NR Vel. W.$\dotfill &Non-rotating line width (m/s)\dotfill &$2400_{-1600}^{+2100}$&$2400_{-1600}^{+2200}$\\
\hline
 Planet Parameters & & & \\
                                  ~~~$P$\dotfill &Period (days)\dotfill &$3.4741085\pm0.0000019$&$3.4741085\pm0.0000019$\\
                           ~~~$a$\dotfill &Semi-major axis (AU)\dotfill &$0.0542_{-0.0021}^{+0.0014}$&$0.0556\pm0.0020$\\
                                 ~~~$M_{P}$\dotfill &3$\sigma$ Mass Limit (\mj)\dotfill &$<3.518$&$<4.165$\\
                               ~~~$R_{P}$\dotfill &Radius (\rj)\dotfill &$1.735_{-0.075}^{+0.070}$&$1.783_{-0.074}^{+0.075}$\\
                           ~~~$\rho_{P}$\dotfill &3$\sigma$ Limit Density (cgs)\dotfill &$<0.840$&$<0.925$\\
                      ~~~$\log{g_{P}}$\dotfill &3$\sigma$ Surface gravity\dotfill &$<3.460$&$<3.511$\\
               ~~~$T_{eq}$\dotfill &Equilibrium temperature (K)\dotfill &$2261\pm73$&$2252_{-79}^{+74}$\\
                           ~~~$\Theta$\dotfill &Safronov number\dotfill &$0.0049_{-0.0040}^{+0.024}$&$0.0045_{-0.0037}^{+0.022}$\\
                   ~~~$\fave$\dotfill &Incident flux (\fluxcgs)\dotfill &$5.93_{-0.73}^{+0.81}$&$5.84_{-0.78}^{+0.80}$\\
\hline
 Radial Velocity Parameters & & & \\
       ~~~$T_C$\dotfill &Time of inferior conjunction (\bjdtdb)\dotfill &$2457485.74965\pm0.00020$&$2457485.74965\pm0.00020$\\
                            ~~~$K$\dotfill &3$\sigma$ RV semi-amplitude (m/s)\dotfill &$<322.51$&$<360.33$\\
                    ~~~$M_P\sin{i}$\dotfill &3$\sigma$ Minimum mass (\mj)\dotfill &$<3.510$&$<4.157$\\
                           ~~~$M_{P}/M_{*}$\dotfill &3$\sigma$ Mass ratio\dotfill &$<0.001925$&$<0.002108$\\
                       ~~~$u$\dotfill &RM linear limb darkening\dotfill &$0.532_{-0.014}^{+0.011}$&$0.533_{-0.015}^{+0.012}$\\
                                ~~~$\gamma_{TRES}$\dotfill &m/s\dotfill &$246_{-96}^{+95}$&$245_{-95}^{+97}$\\
\hline
Linear Ephemeris & &  &            \\
from Follow-up & &  &            \\
Transits: & &  &            \\
                                 ~~~$P_{Trans}$\dotfill &Period (days)\dotfill & $3.4741070\pm0.0000019$&---\\
       ~~~$T_0$\dotfill &Linear ephemeris from transits (\bjdtdb)\dotfill & $2457503.120049\pm0.000190$ &---\\
 \hline
 \hline
 \hline
 \end{tabular}
\begin{flushleft}
 \footnotesize \textbf{\textsc{NOTES}} \\
  \vspace{.1in}
  \footnotesize 3$\sigma$ limits reported for KELT-20b's mass and parameters dependent on mass. 
  \footnotesize The gamma velocity reported here uses an arbitrary zero point for the multi-order relative velocities. The absolute gamma velocity based on the Mg b order analysis is 23.8 +/-0.3 km/s.
 \end{flushleft}
\end{table*}

\begin{table*}
\scriptsize
 \centering
\setlength\tabcolsep{1.5pt}
\caption{Median values and 68\% confidence intervals for the physical and orbital parameters for the KELT-20 System}
  \label{tbl:KELT-17b_part2}
  \begin{tabular}{lccccc}
  \hline
  \hline
  Parameter & Description (Units) & \textbf{Adopted Value} & Value \\
  & & \textbf{(YY circular)} & (Torres circular) \\
 \hline
 \hline
 Primary Transit & & & \\
~~~$R_{P}/R_{*}$\dotfill &Radius of the planet in stellar radii\dotfill &$0.11426\pm0.00062$&$0.11418\pm0.00063$\\
           ~~~$a/R_*$\dotfill &Semi-major axis in stellar radii\dotfill &$7.44_{-0.13}^{+0.14}$&$7.46_{-0.13}^{+0.14}$\\
                          ~~~$i$\dotfill &Inclination (degrees)\dotfill &$86.15_{-0.27}^{+0.28}$&$86.18_{-0.28}^{+0.29}$\\
                               ~~~$b$\dotfill &Impact parameter\dotfill &$0.500_{-0.029}^{+0.026}$&$0.496_{-0.029}^{+0.027}$\\
                             ~~~$\delta$\dotfill &Transit depth\dotfill &$0.01306\pm0.00014$&$0.01304\pm0.00014$\\
                    ~~~$T_{FWHM}$\dotfill &FWHM duration (days)\dotfill &$0.12897\pm0.00048$&$0.12900_{-0.00048}^{+0.00049}$\\
              ~~~$\tau$\dotfill &Ingress/egress duration (days)\dotfill &$0.01985_{-0.00079}^{+0.00082}$&$0.01974_{-0.00080}^{+0.00082}$\\
                     ~~~$T_{14}$\dotfill &Total duration (days)\dotfill &$0.14882_{-0.00090}^{+0.00092}$&$0.14874_{-0.00089}^{+0.00091}$\\
   ~~~$P_{T}$\dotfill &A priori non-grazing transit probability\dotfill &$0.1191\pm0.0021$&$0.1188\pm0.0021$\\
             ~~~$P_{T,G}$\dotfill &A priori transit probability\dotfill &$0.1498_{-0.0027}^{+0.0028}$&$0.1494\pm0.0028$\\
                ~~~$u_{1Sloang}$\dotfill &Linear Limb-darkening\dotfill &$0.3424_{-0.018}^{+0.0090}$&$0.3397_{-0.017}^{+0.0087}$\\
             ~~~$u_{2Sloang}$\dotfill &Quadratic Limb-darkening\dotfill &$0.3362_{-0.0038}^{+0.0073}$&$0.3420_{-0.0083}^{+0.0091}$\\
                ~~~$u_{1Sloani}$\dotfill &Linear Limb-darkening\dotfill &$0.1923_{-0.0084}^{+0.011}$&$0.186_{-0.010}^{+0.012}$\\
             ~~~$u_{2Sloani}$\dotfill &Quadratic Limb-darkening\dotfill &$0.2441_{-0.0063}^{+0.010}$&$0.253_{-0.013}^{+0.026}$\\
                ~~~$u_{1Sloanz}$\dotfill &Linear Limb-darkening\dotfill &$0.1229_{-0.0063}^{+0.0097}$&$0.1179_{-0.0069}^{+0.010}$\\
             ~~~$u_{2Sloanz}$\dotfill &Quadratic Limb-darkening\dotfill &$0.2390_{-0.0080}^{+0.0098}$&$0.246_{-0.012}^{+0.023}$\\
                     ~~~$u_{1V}$\dotfill &Linear Limb-darkening\dotfill &$0.300_{-0.015}^{+0.011}$&$0.295_{-0.015}^{+0.010}$\\
                  ~~~$u_{2V}$\dotfill &Quadratic Limb-darkening\dotfill &$0.3096_{-0.0036}^{+0.0072}$&$0.3171_{-0.0098}^{+0.018}$\\
\hline
Secondary Eclipse & & & \\
                  ~~~$T_{S}$\dotfill &Time of eclipse (\bjdtdb)\dotfill &$2457484.01259\pm0.00020$&$2457484.01260\pm0.00020$\\
\hline
\hline
\end{tabular}
\end{table*}

\subsection{Transit Timing Variation Analysis}
\label{sec:TTVs}
We analyzed the fiducial global model transit center times of all followup light curves (see Table \ref{tab:TTVs}) to search for transit timing variations (TTVs) in the KELT-20 system. Before running the global models, we confirm that all photometric time stamps are in \bjdtdb\ format \citep{Eastman:2010}. To ensure the accuracy of the time stamps, follow-up observers provision telescope control computers to synchronize to a standard clock (such as the atomic clock in Boulder, CO). This synchronization is normally done periodically throughout the observing session. To assess the TTV for each light curve, we find the best linear fit to the transit center times. The resulting linear ephemeris has a reference transit center time of $T_{0}=2457503.120049\pm0.000190$ (\bjdtdb) and a period of $3.4741070\pm0.00000186$ days, and has a $\chi^2$ of 60.8 with 11 degrees of freedom.  We note that the large $\sim 9$ minute TTV in the GCO data (Table \ref{tab:TTVs}) is likely the result of the partial transit coverage and systematics in the light curve (see Figure \ref{fig:All_light curve}). The largest scatter in the other light curves occurs on epoch 109 (see Table \ref{tab:TTVs}) where the transit was simultaneously observed by four telescopes. Using that scatter as the limit of our TTV sensitively threshold, we find no evidence for astrophysical TTVs in our data. We therefore adopt the linear ephemeris specified above as the best predictor of future transit times from our data.

\begin{table}
\centering
 \caption{Transit times from KELT-20 Photometric Observations\MakeLowercase{b}.}
 \label{tab:TTVs}
 \begin{tabular}{r@{\hspace{12pt}} l r r r c}
    \hline
    \hline
    \multicolumn{1}{c}{Epoch} & \multicolumn{1}{c}{$T_\textrm{C}$} 	& \multicolumn{1}{l}{$\sigma_{T_\textrm{C}}$} 	& \multicolumn{1}{c}{O-C} &  \multicolumn{1}{c}{O-C} 			& Telescope \\
	    & \multicolumn{1}{c}{(\bjdtdb)} 	& \multicolumn{1}{c}{(s)}			& \multicolumn{1}{c}{(s)} &  \multicolumn{1}{c}{($\sigma_{T_\textrm{C}}$)} 	& \\
    \hline
-174  & 2456898.624275  &   43   &  -99.14  &  -2.27 & PvdK \\
-166  & 2456926.424578  &  180   &  544.24  &  3.02 & GCO \\
 -58  & 2457301.621915  &   74   &    6.46  &   0.09 & WCO \\
   8  & 2457530.913718  &   56   &   70.20  &   1.23 & DEMONEXT \\
  12  & 2457544.810920  &   44   &  137.06  &   3.08 & DEMONEXT \\
  14  & 2457551.756911  &   55   &  -55.02  &  -1.00 & DEMONEXT \\
  56  & 2457697.671922  &   62   &  162.27  &   2.60 & MINERVA \\
 109  & 2457881.799595  &   48   &  162.22  &   3.37 & PvdK\\
 109  & 2457881.796557  &   49   &  -100.26  &  -2.01& MORC\\
 109  & 2457881.796903  &   55   &  -70.37  &  -1.28 & CDK20N\\
 109  & 2457881.795676  &   75   & -176.38  &  -2.33 & WCO\\
 111  & 2457888.745551  &   51   &  -32.88  &  -0.64 & WCO \\
 119  & 2457916.537500  &   50   & -111.28  &  -2.21 & CROW \\
    \hline
    \hline
 \end{tabular}
  \begin{flushleft}
  \footnotesize{Epochs are given in orbital periods relative to the value of the inferior conjunction time from the global fit.}
\end{flushleft}
\end{table} 

\begin{figure}[!ht]
\includegraphics[width=1\linewidth]{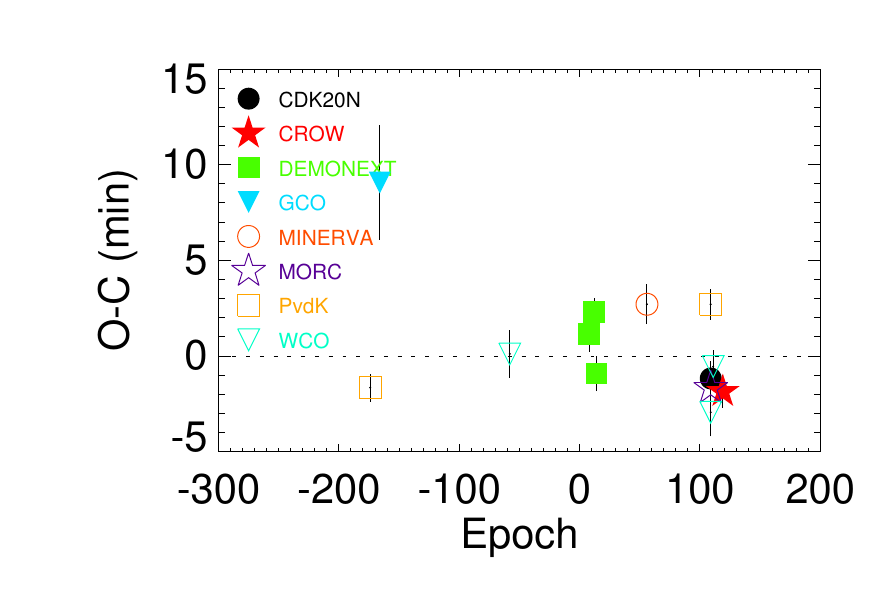}
\caption{The transit time residuals for KELT-20b using the inferior conjunction time from the global fit to define the epoch. The data are listed in Table \ref{tab:TTVs}.}
\label{fig:TTVs}
\end{figure}

\subsection{Doppler Tomographic Characterization}
\label{sec:SpinOrbit}
\begin{figure}
\vspace{0.1in}
\centering
\includegraphics[width=0.95\linewidth, trim = 0 0 0 0]{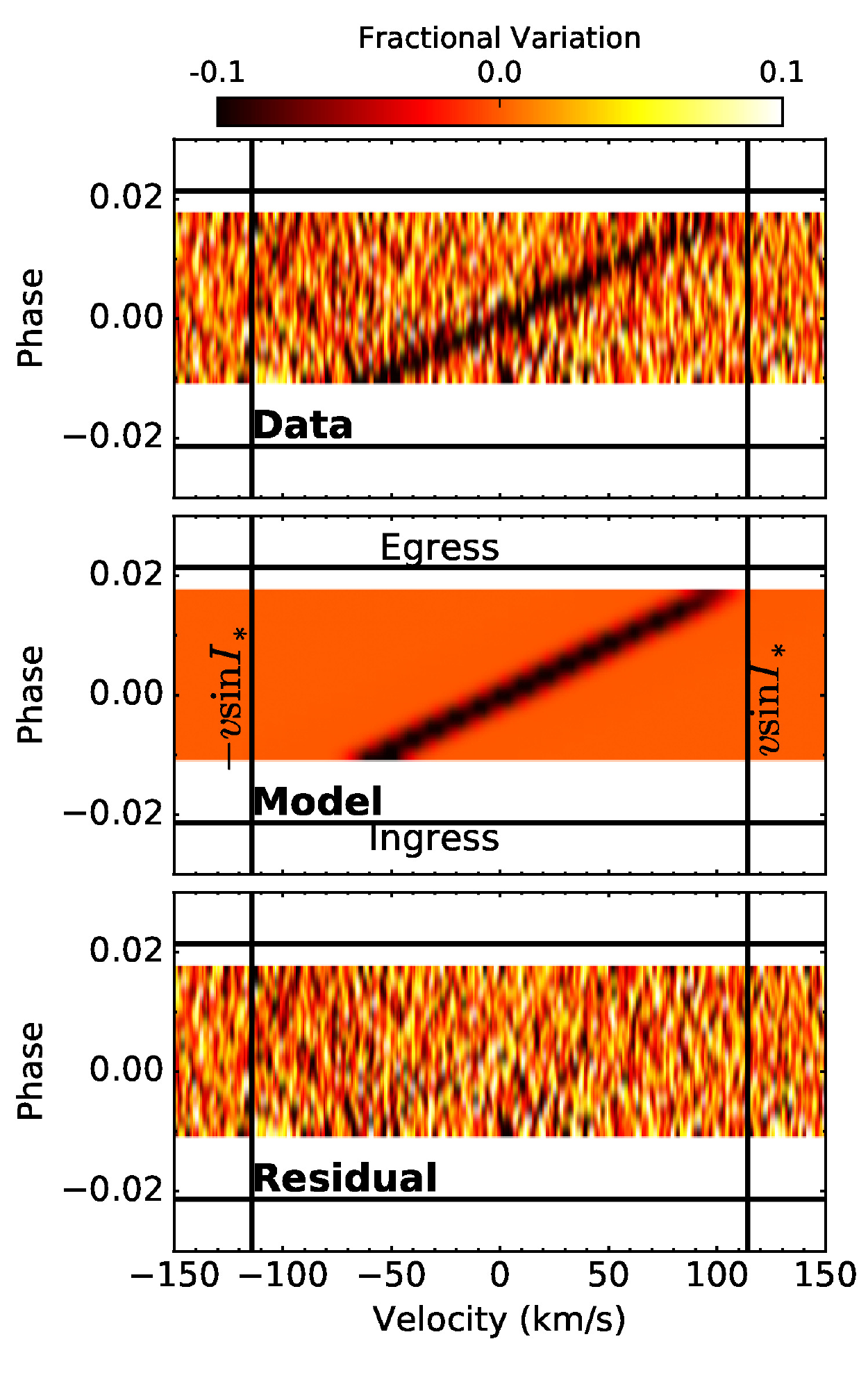}
\caption{\footnotesize The Doppler tomographic transit of KELT-20b, as observed by TRES on UT 2017 April 24. The top panel shows the residuals of the spectroscopic broadening kernels. The temporal axis for the spectral observations is arranged vertically, the velocity axis horizontally. The shadow cast by the planet on the rapidly rotating host star is seen moving across the star, in a spin-orbit aligned geometry, as the dark trail. The best fit model, derived in Section~\ref{sec:GlobalFit}, is shown in the middle panel. The vertical lines mark the boundaries of the stellar rotational profile in terms of $v\sin I_*$. The transit duration is marked with horizontal lines indicating the ingress and egress times. The bottom panel shows the residuals after the model is subtracted.}
\label{fig:doppler}
\end{figure}

\begin{figure}
\vspace{0.1in}
\centering
\includegraphics[width=0.95\linewidth, trim = 0 0 0 0]{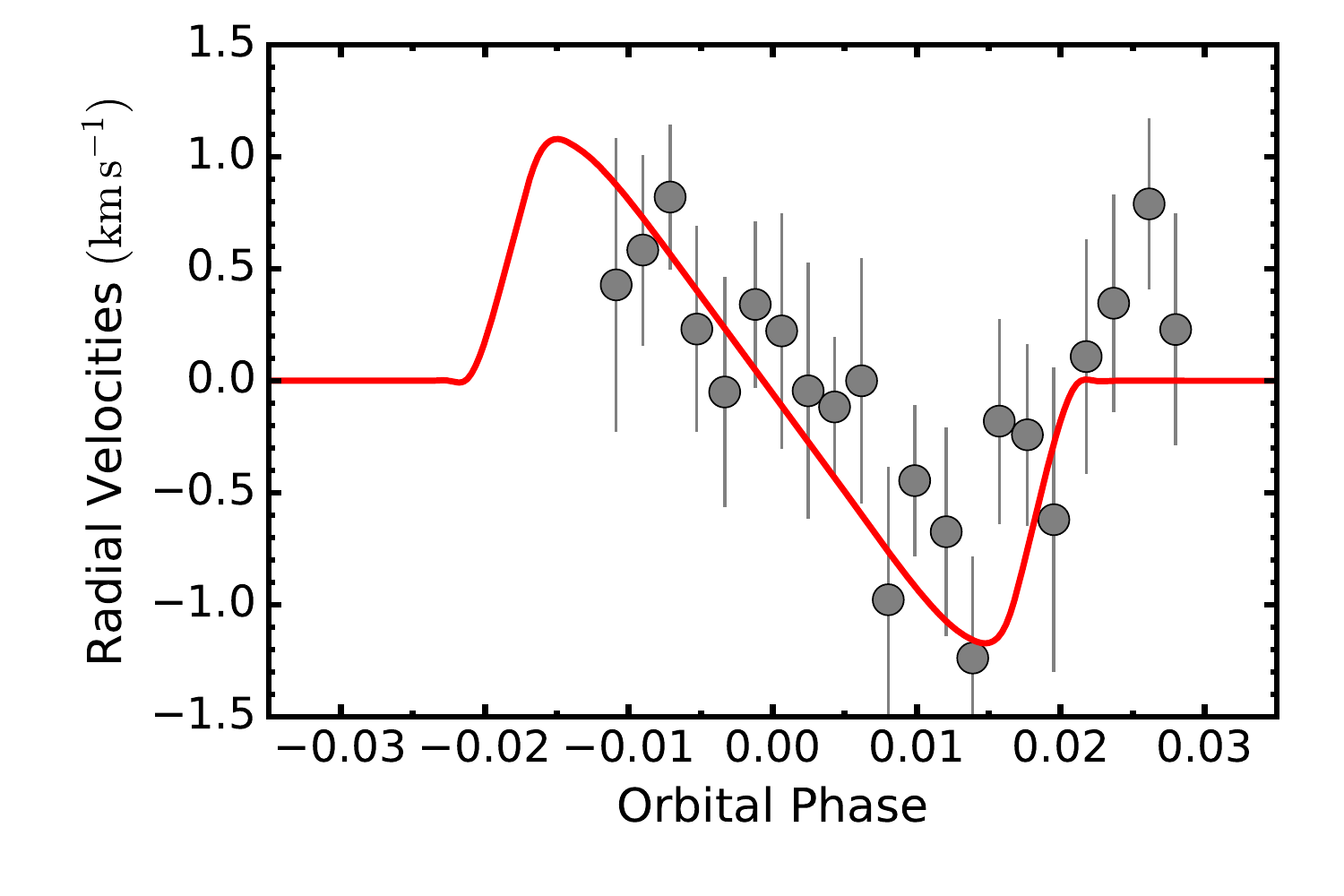}
\caption{\footnotesize The Rossiter-McLaughlin effect was also detected from the same dataset. We plot here the TRES multi-order radial velocities against the expected Rossiter-McLaughlin model, based on the best fit geometry from our global analysis. The Rossiter-McLaughlin signal is modelled using the \emph{ARoME} library \citep{Boue:2013}.
We show these data simply to confirm the consistency with the Doppler  tomographic modelling; the in-transit velocities were not incorporated in the global modelling to avoid double-counting this information. }
\label{fig:RM}
\end{figure}

We obtained 21 in-transit spectroscopic observations of KELT-20b with TRES on 2017-04-24. These observations were made and processed as per \citet{Zhou:2016}. For each spectrum, we derive a rotational profile via a least-squares deconvolution against a non-rotating template spectrum, as per the techniques described in \citet{Donati1997} and \citet{CollierCameron2010}. We create a median-combined rotational profile that averages out the transit signal. This median-combined rotational profile is then subtracted from each individual exposure, revealing the dark shadow of the planet transiting across the star (Figure~\ref{fig:doppler}). These line profile residuals are modeled in the global analysis in Section~\ref{sec:GlobalFit} as described in \citet{Gaudi:2017}. We adopt linear limb darkening coefficients from \citet{Claret:2004} for the $V$ band in the Doppler tomographic modelling. By modeling the rotational broadening profiles, we also measured rotational broadening parameters $v\sin I_*$ of $114.92 \pm 4.24\,\kms$ and a macroturbulence velocity of $6.08_{-2.03}^{+4.44}\,\kms$. These were adopted as Gaussian priors in the global analysis in Section~\ref{sec:GlobalFit}. In addition, we also checked the transit Doppler tomography result by deriving multi-order radial velocities for the same dataset. These velocities also clearly show the Rossiter-McLaughlin effect \citep{Rossiter:1924,McLaughlin:1924} consistent with the spin-orbit angle derived from the global analysis (see Figure~\ref{fig:RM}). 

\subsection{False-Positive Analysis}
\label{sec:False-Positives}
Despite the unusual nature of this system, and the lack of a definitive measurement of the companion mass, we are confident that this system is truly a hot Jupiter transiting an early A star.  The evidence for this comes from several sources which we will briefly review, however we invite the reader to review papers by \citet{Bieryla:2015,Zhou:2016,Zhou:2017} and \citet{Hartman:2015} for a more detailed explanation.  Of course, the first system to have been validated in this way was WASP-33b \citep{CollierCameron2010}.

The Doppler tomographic observation eliminates the possibility of a blended eclipsing binary causing the transit signal. The line profile derived from the least-squares deconvolution shows a lack of spectroscopic companions blended with KELT-20. The spectroscopic transit is seen crossing the entirety of the rapidly rotating target star's line profile, confirming that it is indeed orbiting KELT-20. The summed flux underneath the Doppler tomographic shadow and the distance of closest approach of the shadow from the zero velocity at the center of the predicted transit time is consistent with both the photometric transit depth and impact parameter, suggesting that the photometric transit is not diluted by background stars, and is fully consistent with the spectroscopic transit. 

Adaptive optics observations (Section~\ref{sec:AO}) also eliminate blended stars with $\Delta K<7.5$ and $>0.6\arcsec$ of KELT-20, consistent with the lack of blending in the spectroscopic analysis. 

Finally, the planetary nature of KELT-20b is confirmed by the TRES radial velocity measurements, which constrain the mass the companion to be $\la 3.5 \,M_\mathrm{jup}$ at $3\sigma$ significance. This eliminates the possibility that the transiting companion is a stellar or brown-dwarf-mass object. As such, KELT-20b is confirmed as a planetary-mass companion transiting the rapidly rotating A star HD~185603.

Thus we conclude that all the available evidence suggests that the most plausible interpretation is that KELT-20b is a Jupiter-size planet transiting an early A-star with a projected spin-orbit alignment that is (perhaps surprisingly) well-aligned (see \ref{sec:spin-orbit-alignment}).

\section{Discussion}
\label{sec:Discussion}

The KELT-20 system represents one of the most extreme transiting hot Jupiter systems, and indeed one of the  most extreme transiting exoplanet systems, yet discovered, by several measures.  The host star is both exceptionally bright ($V\sim 7.6$), and exceptionally hot ($\teff \simeq 8700$K).  It is only the sixth A star known to host a transiting giant companion.  The planet itself is on a relatively short period orbit of $P\simeq 3.5$~days, and thus receives an extreme amount of stellar insolation, resulting in an estimated equilibrium temperature of $\sim 2250$~K.  Because its host is an A star, it also receives a higher amount of high-energy radiation than the majority of known transiting planet systems, which may lead to significant atmospheric ablation \citep{Murray-Clay:2009}. 

There are two additional notable facts about the KELT-20 system.  First, the host star appears to be quite young, with a main-sequence age of $\la 600$~Myr (see Sec.~\ref{fig:hrd}). Whether or not this places interesting constraints on the migration timescale of its hot Jupiter should be considered.  Second, and perhaps relatedly, the planet's orbit normal appears to be well-aligned with the spin axis of the star (see Sec.~\ref{sec:spin-orbit-alignment}), which is generally atypical for hot Jupiters orbiting hot stars \citep{Winn:2010,Schlaufman:2010}.  

\subsection{Prospects for Characterization}
\label{Characterization}

In many ways, KELT-20b appears to be quite similar to KELT-9b \citep{Gaudi:2017}, albeit orbiting a slightly cooler and less massive star at a somewhat longer ($\sim 2.3$ times) period.  However, the fact that KELT-20 is nearly as bright as KELT-9 nevertheless makes the prospect for characterization of the system nearly as promising as for KELT-9b. 

\begin{figure}
\vspace{0.1in}
\includegraphics[width=1\linewidth]{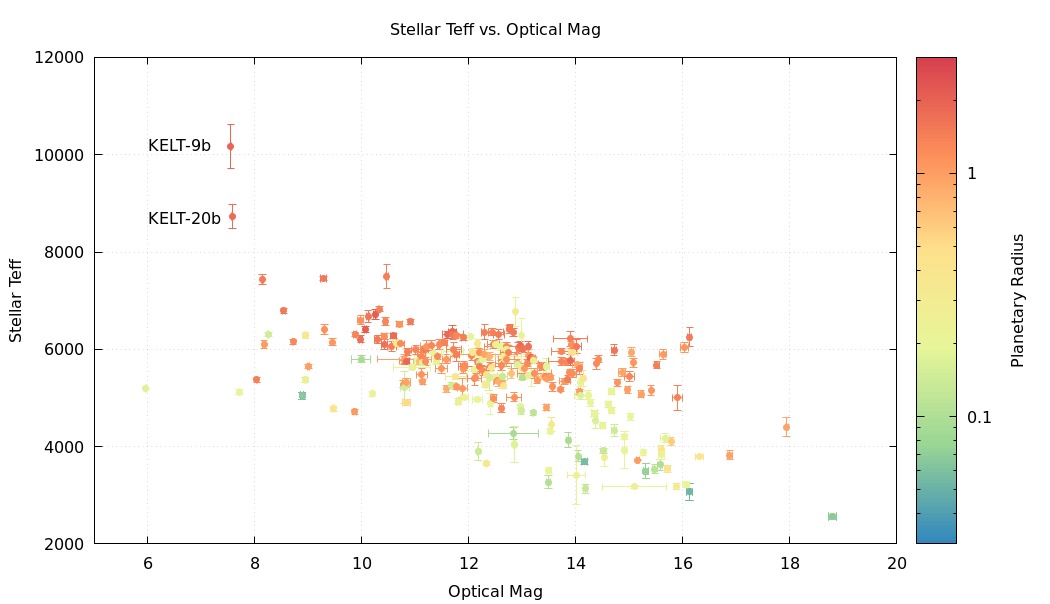}
\caption{The population of transiting exoplanets based on the host star's optical magnitude and effective temperature (\teff), with colors indicating the radius of the planet in \rj. The bulk of these data come from the NASA Exoplanet Database \footnote{https://exoplanetarchive.ipac.caltech.edu}, with the addition of KELT-20b to this data set. The figure was plotted using Filtergraph \citep{Burger:2013}, and the data set for the plot can be found here: https://filtergraph.com/KELT20b\_StellarComparison.}
\label{fig:Teff_Mag}
\end{figure}

Figure \ref{fig:Teff_Mag} shows the host star effective temperature versus the $V$-band magnitude for known transiting planets.  Together with 55 Cancri \citep{Winn:2011,Demory:2011b}, KELT-9b and KELT-20b are the three brightest (in $V$) transiting planet hosts known, while KELT-9b and KELT-20b are the two brightest hosts of transiting hot Jupiters, which are considerably more amenable to detailed follow-up.

\begin{figure}[!ht]
\includegraphics[width=1\linewidth]{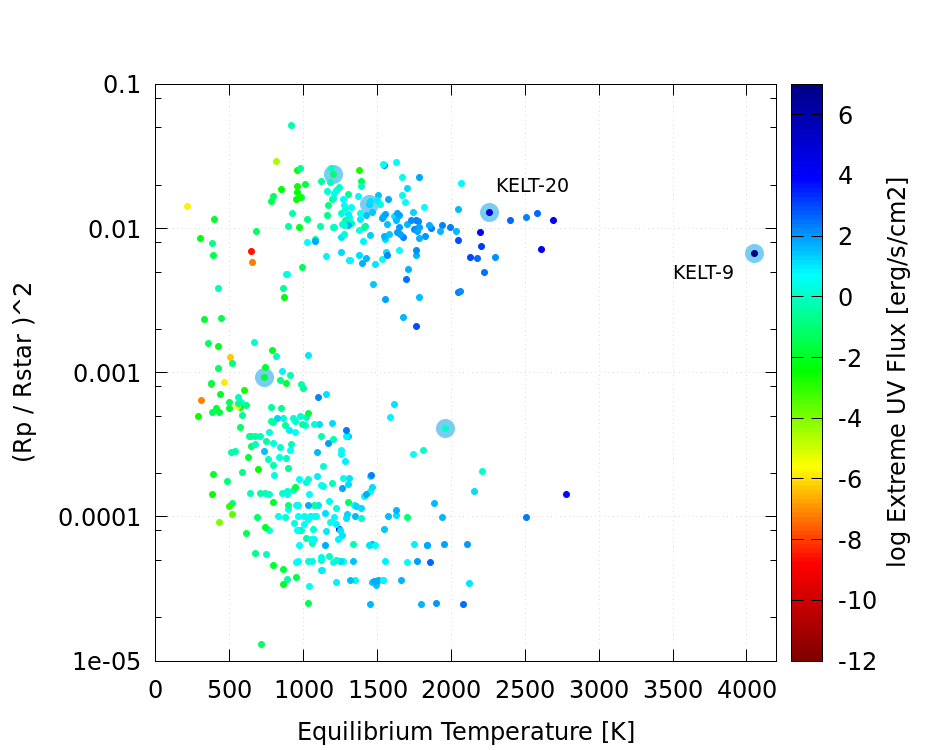}
\caption{
Depth of the transit signal, $(R_P/\rstar)^2$, versus equilibrium temperature assuming zero albedo and complete heat redistribution for known transiting planets with $V<13$.  Those with $V<8$ are shown with large symbols.  The points are color coded by the amount of incident extreme ultraviolet ($\lambda \le 91.2$~nanometers) flux the planet receives from its parent star. In the case of the stars with $V<8$ the color in the middle of the symbol represents this value.}
\label{fig:RpRstar}
\end{figure}

Figure \ref{fig:RpRstar} shows the primary transit depth, $\delta=(R_P/\rstar)^2$, versus predicted planetary equilibrium temperature \teq\ (assuming zero albedo and complete heat redistribution) for planets with host stars $V<13$, color coded by the amount of UV flux the planet receives. Although KELT-20b's predicted equilibrium temperature is not nearly as high as KELT-9b, it is nevertheless one of the hottest dozen or so known hot Jupiters.  Furthermore, its transit depth is nearly twice that of KELT-9b. Although we only have an upper limit on the mass of KELT-20b, our $3\sigma$ upper limit on the surface gravity \loggplanet\ is $\sim 3.5$ (cgs).  We can therefore predict that the magnitude of the thermal emission spectrum, transmission spectrum, and phase curve should all be easily detectable with {\it Spitzer}, the {\it Hubble Space Telescope (HST)}, and eventually the {\it James Webb Space Telescope}.  Indeed, the planet is sufficiently hot that secondary eclipse measurements should be possible from ground-based instruments.  We also expect that, should the atmosphere be significantly ablated by the high UV flux incident on the planet, this may be detectable via {\it HST}.

\subsection{Comparison to KELT-9 and other A star hosts of giant transiting planets}
\label{sec:Comparison}

\begin{figure}[!ht]
\includegraphics[width=0.75\linewidth, angle=90, trim = 0 0 0 0]{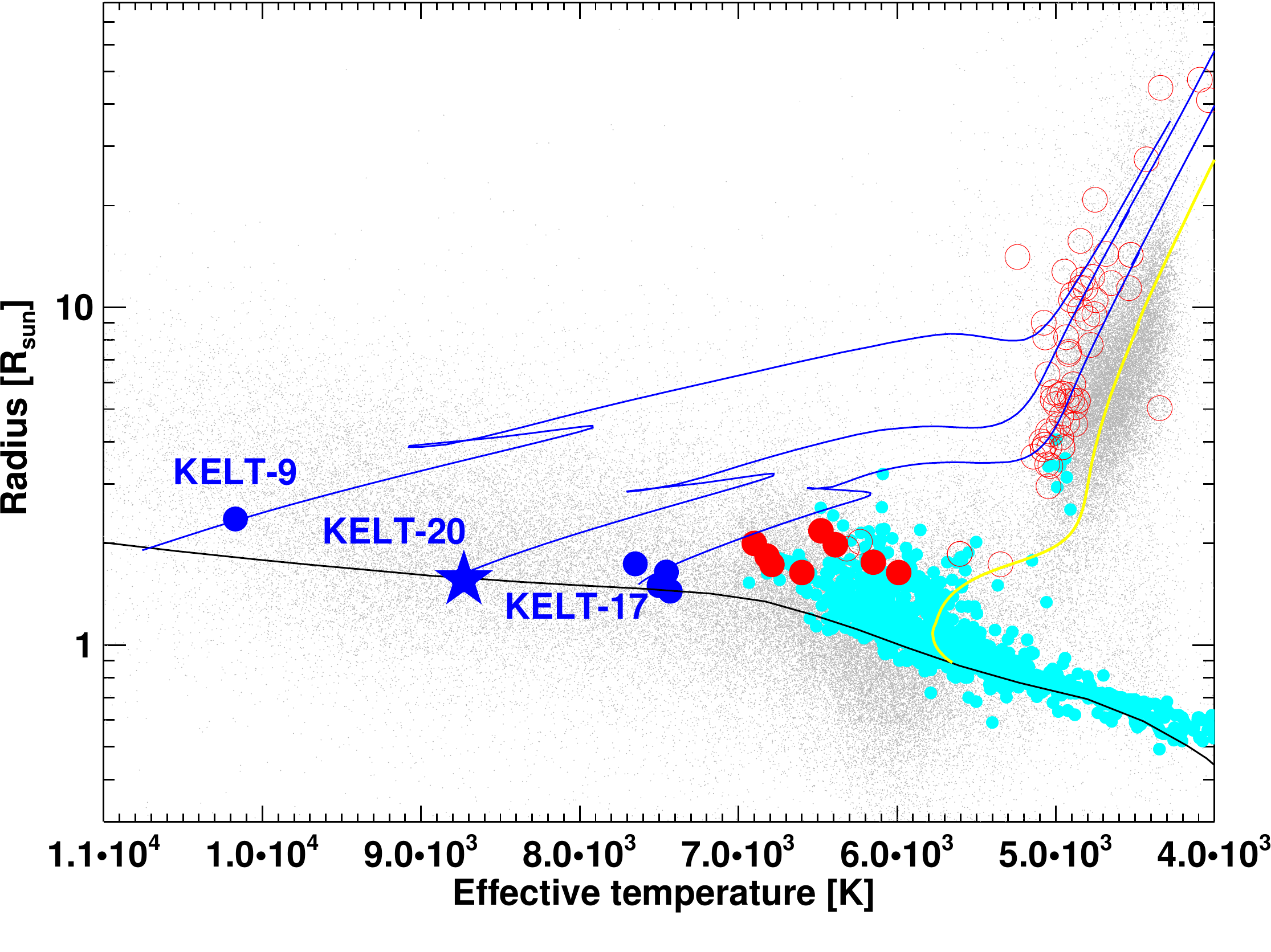}
\caption{
Radius versus effective temperature of hosts of known planets detected by the radial-velocity (open circles) and transit methods (filled circles), as well as nearby stars in the Hipparcos catalog for reference (grey points).  Only planet hosts with $V\le 10$ are shown for clarity. The cyan symbols are low-mass planet hosts with $M < 1.4~M_\odot$, red symbols indicate massive planet hosts with $M_* \ge 1.4~M_\odot$. 
The yellow line shows the evolutionary trajectory for a solar analog ($M_*=M_\odot$ and solar metallicity), whereas the blue tracks shows the evolutionary trajectories
for KELT-9, KELT-20, and KELT-17.  The other three blue circles are (from left to right) Kepler-13, HAT-P-57, and WASP-33. We also show the zero-age main sequence (ZAMS) for solar-metallicity stars from the YY isochrones (black curve). 
}
\label{fig:HRD_context}
\end{figure}

With a sample of six A star hosts of transiting gas giants now known, it starts to become possible to consider and compare the ensemble properties of such systems.  Figure \ref{fig:HRD_context} shows one such comparison, namely the location and expected future evolution of these hosts on a $\rstar$ versus $\teff$ (modified Hertzsprung-Russell) diagram. We show the evolutionary tracks based on the YY isochrones for KELT-9 ($\mstar \simeq 2.52~M_\odot$), KELT-20 ($\mstar \simeq 1.76$\msun), and KELT-17 ($\mstar \simeq 1.63$\msun), all assuming solar metallicity.  The other three blue circles are (from left to right) Kepler-13 A ($\teff \simeq 7650$K), HAT-P-57 ($\teff \simeq 7500$K), and WASP-33 ($\teff \simeq 7430$K), all of which have quite similar $\teff$ as KELT-17, and radii and masses that differ by only $\sim 20\%$.  

We note that while KELT-9, KELT-17, and Kepler-13 are somewhat evolved from the ZAMS, KELT-20, and to a lesser extent HAT-P-57 and WASP-33, appear to be on (or perhaps even slightly below) the ZAMS, indicating that they are young, or (less likely) have subsolar metallicity.

\subsubsection{Spin-Orbit Alignment}
\label{sec:spin-orbit-alignment}
Doppler tomographic observations allow the measurement of the spin-orbit misalignment ($\lambda$). This, however, is merely the {\it sky-projected} angle between the stellar spin and planetary orbital angular momentum vectors. Measurement of the full three-dimensional spin-orbit angle ($\psi$) requires knowledge of the inclination of the stellar rotation axis with respect to the line of sight ($I_*$), which is typically difficult to measure. We do not have such a measurement of this angle for KELT-20, and so cannot directly calculate $\psi$.

We can, however, set limits upon $I_*$, and thus upon $\psi$. Following \cite{Iorio:2011}, we can limit $I_*$ by requiring that the star be rotating at less than break-up velocity. Using our measured stellar and planetary parameters, we obtain a $1\sigma$ limit of $24.4^{\circ}<I_*<155.6^{\circ}$. Together with our measured values of $\lambda$ and $i$, this implies $1.3^{\circ}<\psi<69.8^{\circ}$ (again at 1$\sigma$). 

Although the planetary orbit is well-aligned if $I_*$ is close to $90^{\circ}$ (i.e., the stellar rotation axis is close to perpendicular to the line of sight), in which case $\psi\sim\lambda$, it may still be substantially misaligned if we are viewing the star closer to pole-on. KELT-20 has a projected rotational velocity of $v\sin I_*=115.9 \pm 3.4$ km s$^{-1}$, which is slightly lower than the median deprojected rotational velocity of 131 km s$^{-1}$ found by \cite{Royer:2007} for A2-A3 main sequence stars. This suggests that KELT-20 is plausibly close to equator-on and approximately aligned. However, we cannot exclude the possibility that KELT-20 is rotating faster than the median for similar stars and the orbit is misaligned. 

A measurement or constraint on $I_*$ may be possible in the future via several methods. First, the detection of rotational modulation would constrain the rotation period and thus $I_*$, however, this is unlikely and difficult for a hot, likely inactive A star like KELT-20. An asteroseismic measurement of the rotation rate is possible by measuring the rotational splitting of the modes. However, there is no evidence that KELT-20 is pulsating, and thus this would require long-time-baseline, very high precision space-based photometry. It may be possible to measure $I_*$ using very high precision light curves affected by gravity darkening \citep{Barnes:2009}, or by measuring the nodal precession of the planet if it is not aligned \citep{Johnson:2015,Iorio:2016}. Even in the most optimistic case, however, the precession rate will be $d\Omega/dt<0.03^{\circ}$ yr$^{-1}$. This is at least an order of magnitude smaller than that measured for WASP-33b by \cite{Johnson:2015}, and would take several decades to give rise to a detectable change in $\lambda$ or $b$.

\begin{figure}
\vspace{0.1in}
\centering
\includegraphics[width=0.95\linewidth, trim = 0 0 0 0]{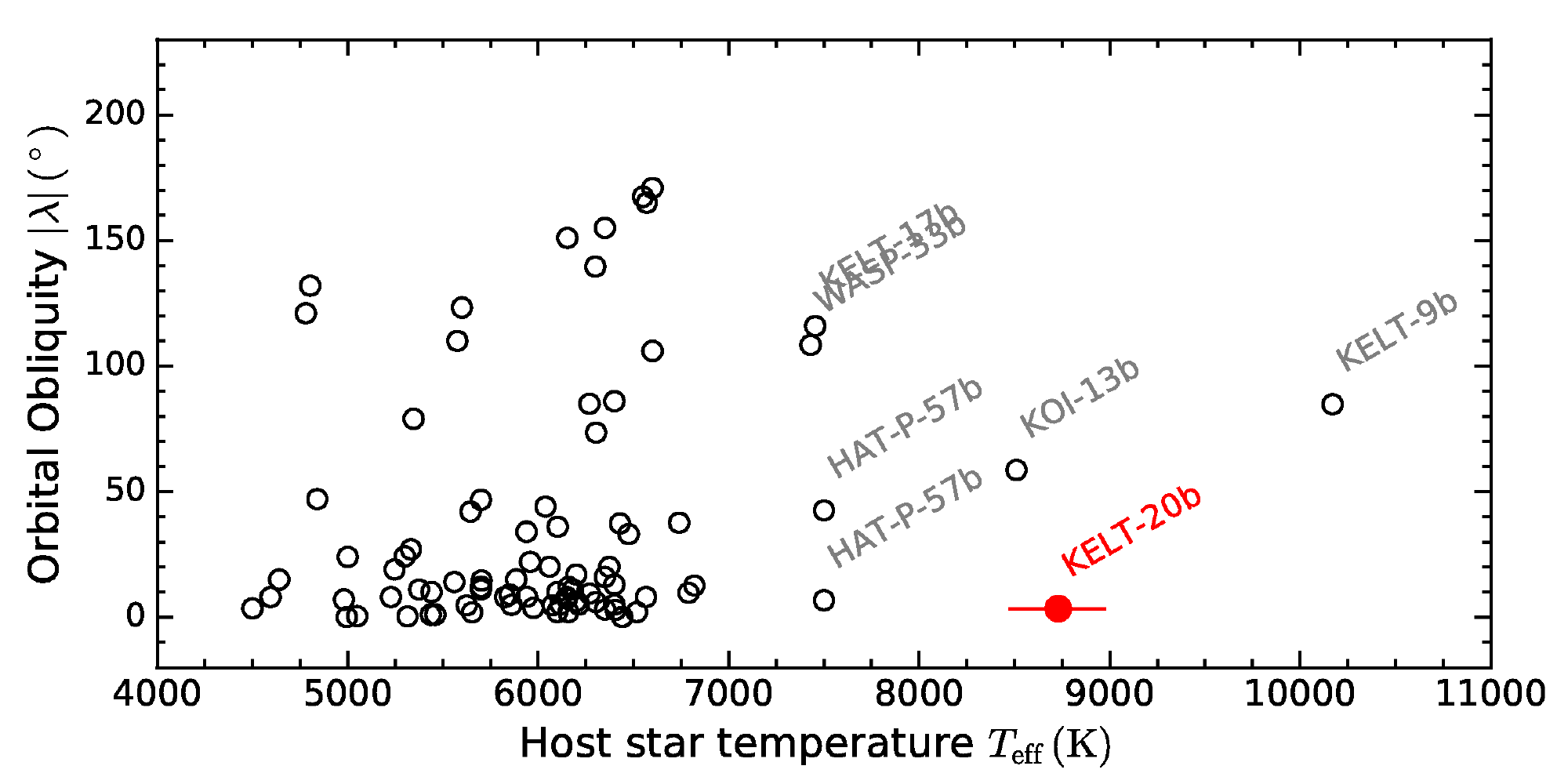}
\caption{\footnotesize Projected spin-orbit angle of all transiting planets measured to date. Planets around host stars with $T_\mathrm{eff} > 7000\,\mathrm{K}$ are labelled. KELT-20b is only the sixth hot Jupiter found around an A-star, and the first of those to be confirmed in projected spin-orbit alignment. Note that two solutions for the projected spin-orbit angle were offered by \citet{Hartman:2016} for HAT-P-57b.}
\label{fig:SpinOrbit}
\end{figure}

Because of its larger mass and therefore more rapid evolution, KELT-20 is likely to be exceptionally young ($<600$~Myr) if it has a near-solar metallicity, as expected.  This may place interesting constraints on the timescale for its migration to its current orbit.  The fact that KELT-20b is one of only two hot Jupiters orbiting A-type stars that could have an aligned orbit\footnote{\citet{Hartman:2015} obtained a bimodal distribution for $\lambda$ for HAT-P-57b, indicating either an aligned orbit or a prograde orbit with a substantial misalignment}, as shown in Figure~\ref{fig:SpinOrbit}, may be particularly interesting in this regard. 

\subsubsection{The Past and Future Evolution of the KELT-20 system}
\label{sec:PastFuture}

We note that KELT-20 is a somewhat unusual system as compared to many hot Jupiters in that the spin period of the star is shorter than the orbital period of the planet.  This implies that tides serve to increase the semimajor axis of the planet, rather than to decrease it.  Furthermore, as the star has essentially no convective envelope, one would expect tides to behave quite differently than in stars with convective envelopes.  Finally, the expected large oblateness of the host star may affect the efficiency and nature of tidal dissipation.  

\begin{figure}
\vspace{0.1in}
\includegraphics[width=1.00\linewidth]{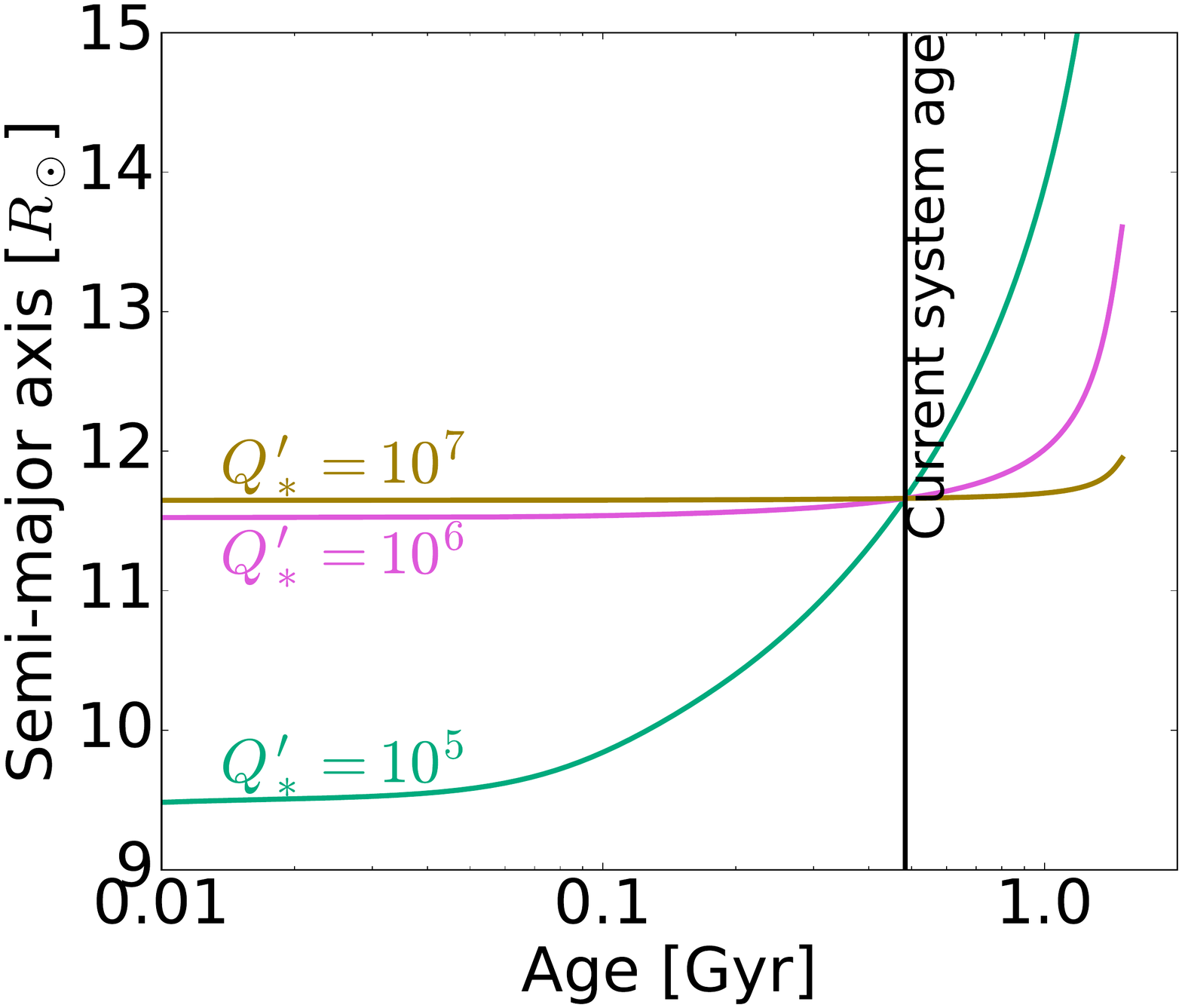}
\includegraphics[width=1.00\linewidth]{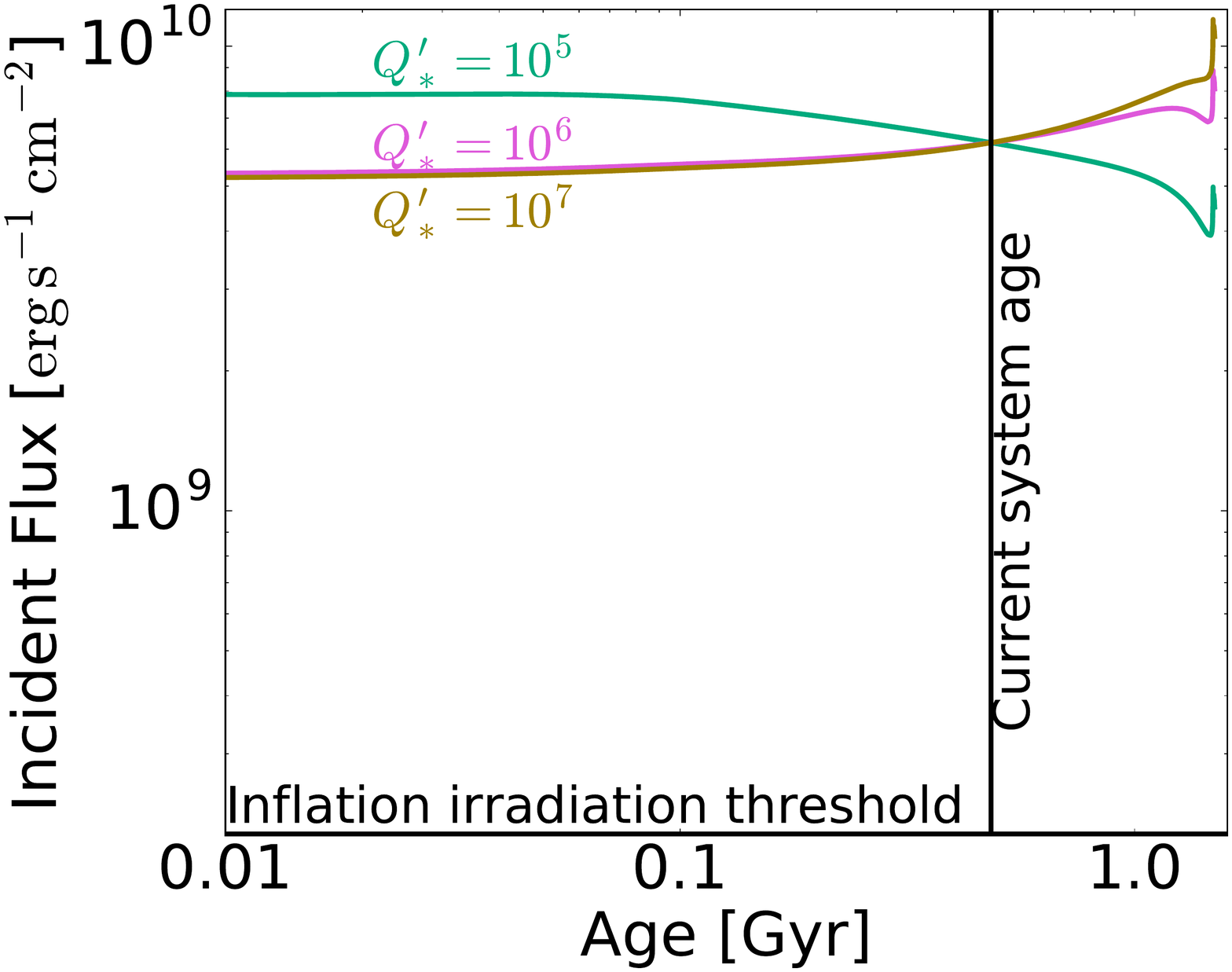}
\caption{(Top) 
Predicted past and future tidal evolution of the semimajor axis of KELT-20b in units of the solar radius as a function of the age of the system.  The current age is assumed to be roughly $480$~Myr.  The evolution is shown under the assumption of a constant tidal phase lag, and for various values of $Q'_\star$, where $1/Q'_\star$ is the product of the phase lag and the stellar Love number. 
(Bottom)  The stellar insolation from the star received by the planet for the same assumptions as above.
}
\label{fig:Past}
\end{figure}

Nevertheless, we proceed to estimate the past and future orbital evolution of the system under tides.  Specifically, we compute the evolution of the semimajor axis in units of the stellar radius, and the evolution of the stellar insolation.

The orbital evolution of KELT-20b was calculated under the assumption of a constant phase lag, including the effect of the changing stellar radius due to stellar evolution, following \citet{Penev:2014}. Due to the poorly constrained efficiency of tidal dissipation in stars, we consider a wide range of dissipation parameters ($Q'_\star = 10^5$, $10^6$ and $10^7$), where $1/Q'_\star$ is the product of the phase lag and the stellar tidal Love number. Given a dissipation parameter, the initial orbital period of the planet was chosen such that the currently observed orbital period is reproduced at an age of 480 Myr. Note that the least dissipative case considered here ($Q'_\star = 10^7$) was chosen simply because it leads to very little orbital evolution, and is in no way physically motivated.

Figure \ref{fig:Past} shows the past and future evolution of the orbit of the planet relative to the stellar radius as a function of the age of the system under these assumptions.  As mentioned above, unlike the majority of hot Jupiter systems, the measured $\vsinistar$ of the host star implies that the stellar spin period is shorter than the orbital period. As a result, the typical picture of a decaying orbit is reversed and the orbit expands over time due to tidal dissipation.  Even under the fairly unrealistic value of $Q'_\star \sim 10^5$, the planet will avoid engulfment by the star until well after it begins to extend up the giant branch.   

Figure \ref{fig:Past} also shows the past and future evolution of stellar incident insolation flux received by the planet. The increase in the planet's orbit due to tides is roughly offset by the increase in the radius of the star due to stellar evolution. KELT-20b was likely always above the empirically-estimated minimum insolation for inflated giant planets \citep{Demory:2011}, which is not suprising given its inferred radius of $R_P\sim 1.6~\rj$. 

Note that at around 1.5~Gyr, the star will cross the Kraft break \citep{Kraft:1967} and begin to develop a deep convective envelope.  However, it is unlikely that the planet will have synchronized its period with that of the star, and so we do not expect this system to evolve into an RS CVn system (c.f.\ \citealt{Siverd:2012}).  KELT-20 will eventually engulf its planet, but not until it has ascended the giant branch.  

\section{Summary}

We have presented the discovery of KELT-20b, currently the third brightest transiting planet system, and the second brightest transiting hot Jupiter system.  The host star is an early A star with an effective temperature of $\teff \simeq 8700$K.  The host is rapidly rotating, with $\vsinistar \sim 116~\kms$. This rapid rotation made confirmation of the planet difficult using radial velocities, and we were only able to obtain an $3\sigma$ upper limit on the mass of the planet of $\sim 3.5~\mj$.  Nevertheless, we confirm the planetary nature of the companion via Doppler tomography, which perhaps surprisingly shows that the orbit normal of the planet is well-aligned with the projected spin-axis of the star.  

The planet has a period of $\sim 3.5$ days, and an equilibrium temperature of $\sim 2250$K, assuming zero albedo and perfect heat redistribution.  With a visual magnitude of $7.6$, an exceptionally high equilibrium temperature, and a likely large scale height, it is an excellent target for detailed follow-up and characterization of a hot Jupiter suffering from extreme stellar irradiation, particularly UV stellar irradiation. 

We infer a surface gravity for the star that is surprisingly large, indicating that the star is either exceptionally young, or (less likely) has a low metallicity compared to solar.  We therefore encourage studies that determine whether or not the likely young age places interesting constraints on the timescale for the planet's migration.  

Finally, with a total of six A-star hosts to transiting gas giants now known, we can begin to compare and contrast the ensemble properties of these systems, and ultimately learn about their origins, as well as their future evolution.

{\bf Note:} During the preparation of this paper, our team became aware of another paper by The Multi-site All-Sky CAmeRA (MASCARA) collaboration \citep{Talens:2017} reporting the discovery of a planetary companion to the host star discussed here, HD 185603 (Talens et al. submitted). While we assume this planetary companion is indeed KELT-20b, no information about the analysis procedure or any results were shared between our groups prior to the submission of both papers.  We would like the thank the MASCARA collaboration for their collegiality and willingness to work with the KELT collaboration to coordinate our announcements of these discoveries simultaneously. 

\section{Acknowledgements}
Work performed by J.E.R. was supported by the Harvard Future Faculty Leaders Postdoctoral fellowship.
D.J.S and B.S.G. were partially supported by NSF CAREER Grant AST-1056524.
Work by S.V.Jr. is supported by the National Science Foundation Graduate Research Fellowship under Grant No. DGE-1343012.
KGS acknowledges partial support from NSF PAARE grant AST-1358862.
This work has made use of NASA's Astrophysics Data System, the Extrasolar Planet Encyclopedia, the NASA Exoplanet Archive, the SIMBAD database operated at CDS, Strasbourg, France, and the VizieR catalogue access tool, CDS, Strasbourg, France.  We make use of Filtergraph, an online data visualization tool developed at Vanderbilt University through the Vanderbilt Initiative in Data-intensive Astrophysics (VIDA).
We also used data products from the Widefield Infrared Survey Explorer, which is a joint project of the University of California, Los Angeles; the Jet Propulsion Laboratory/California Institute of Technology, which is funded by the National Aeronautics and Space Administration; the Two Micron All Sky Survey, which is a joint project of the University of Massachusetts and the Infrared Processing and Analysis Center/California Institute of Technology, funded by the National Aeronautics and Space Administration and the National Science Foundation; and the European Space Agency (ESA) mission {\it Gaia} (\url{http://www.cosmos.esa.int/gaia}), processed by the {\it Gaia} Data Processing and Analysis Consortium (DPAC, \url{http://www.cosmos.esa.int/web/gaia/dpac/consortium}). Funding for the DPAC has been provided by national institutions, in particular the institutions participating in the {\it Gaia} Multilateral Agreement.

\bibliography{ms}{}
\bibliographystyle{apj}

\end{document}